\documentclass[sigconf]{acmart}

\usepackage{booktabs}
\usepackage{longtable}
\usepackage{tabularx}
\usepackage{multirow}
\usepackage{array}
\usepackage{ragged2e}
\usepackage{caption}
\usepackage{threeparttable}
\usepackage{threeparttable} 
\AtBeginDocument{%
  \providecommand\BibTeX{{%
    \normalfont B\kern-0.5em{\scshape i\kern-0.25em b}\kern-0.8em\TeX}}}

\acmYear{2025}
\setcopyright{cc}
\setcctype{by}
\acmConference[ASSETS '25]{The 27th International ACM SIGACCESS Conference
on Computers and Accessibility}{October 26--29, 2025}{Denver, CO, USA}
\acmBooktitle{The 27th International ACM SIGACCESS Conference on Computers
and Accessibility (ASSETS '25), October 26--29, 2025, Denver, CO, USA}
\acmDOI{10.1145/3663547.3746365}
\acmISBN{979-8-4007-0676-9/2025/10}

\usepackage{todonotes}
\usepackage{soul}  % For highlighting text with \hl

\begin{document}

%%
%% The "title" command has an optional parameter,
%% allowing the author to define a "short title" to be used in page headers.
\title[PunchPulse - An Accessible VR Boxing Exergame]{PunchPulse: A Physically Demanding Virtual Reality Boxing Game Designed with, for and by Blind and Low-Vision Players}

%%
%% The "author" command and its associated commands are used to define
%% the authors and their affiliations.
%% Of note is the shared affiliation of the first two authors, and the
%% "authornote" and "authornotemark" commands
%% used to denote shared contribution to the research.

\author{Sanchita S. Kamath}
\orcid{0000-0001-6469-0360}
\affiliation{%
 \department{School of Information Sciences}
  \institution{University of Illinois Urbana-Champaign}
  \city{Champaign}
  \state{Illinois}
  \country{USA}
  \postcode{61820}
  }
\email{ssk11@illinois.edu}

\author{Omar Khan}
\orcid{0009-0005-3209-3525}
\affiliation{%
 \department{Computer Science}
  \institution{University of Illinois Urbana-Champaign}
  \city{Urbana}
  \state{Illinois}
  \country{USA}
  \postcode{61801}
  }
\email{mkhan259@illinois.edu}

\author{Anurag Choudhary}
\orcid{0000-0003-1043-3830}
\affiliation{%
 \department{Computer Science}
  \institution{University of Illinois Urbana-Champaign}
  \city{Urbana}
  \state{Illinois}
  \country{USA}
  \postcode{61820}
  }
\email{anuragc3@illinois.edu}

\author{Jan Meyerhoff-Liang}
\orcid{0000-0001-8752-9684}
\affiliation{%
   \department{ Institute for New Economic Thinking}
  \institution{Oxford Martin School}
  \city{Oxfordshire}
  \country{UK}
  }
\email{Jan.Meyerhoff@inet.ox.ac.uk}

\author{Soyoung Choi}
\orcid{0000-0002-0998-3352}
\affiliation{%
 \department{Department of Kinesiology and Community Health}
  \institution{University of Illinois Urbana-Champaign}
  \city{Urbana}
  \state{Illinois}
  \country{USA}
  \postcode{61820}
  }
\email{soyoung@illinois.edu}

\author{JooYoung Seo}
\orcid{0000-0002-4064-6012}
\affiliation{%
 \department{School of Information Sciences}
  \institution{University of Illinois Urbana-Champaign}
  \city{Champaign}
  \state{Illinois}
  \country{USA}
  \postcode{61820}
  }
\email{jseo1005@illinois.edu}

%%
%% By default, the full list of authors will be used in the page
%% headers. Often, this list is too long, and will overlap
%% other information printed in the page headers. This command allows
%% the author to define a more concise list
%% of authors' names for this purpose.
\renewcommand{\shortauthors}{Kamath et al.}

%%
%% The abstract is a short summary of the work to be presented in the
%% article.
\begin{abstract}
  Blind and low-vision (BLV) individuals experience lower levels of physical activity (PA) due to limited access to engaging, accessible exercise tools. We present \textit{\textbf{PunchPulse}}, an open-source VR boxing exergame (available on \href{https://github.com/xability/punch-pulse}{GitHub}) designed in collaboration with BLV co-designers to support sustained moderate-to-vigorous physical activity (MVPA) through immersive, autonomous gameplay. Our system emphasizes structured progression and multimodal interaction (e.g., heart-rate tracking, audio-haptic feedback) to scaffold engagement and exertion. Over a seven-month, multi-phased study, \textit{\textbf{PunchPulse}} was iteratively refined with three BLV co-designers, informed by two early pilot testers, and evaluated by six additional BLV user-study participants. Data collection included both qualitative (researcher observations, semi-structured interviews) and quantitative (MVPA zones, aid usage, completion times) measures of physical exertion and gameplay performance. The user study revealed that all participants reached moderate MVPA thresholds, with high levels of immersion and engagement observed. This work demonstrates the potential of VR as an inclusive medium for promoting meaningful PA in the BLV community and addresses a critical gap in accessible, intensity-driven exercise interventions.
\end{abstract}

%%
%% The code below is generated by the tool at http://dl.acm.org/ccs.cfm.
%% Please copy and paste the code instead of the example below.
%%


%%
%% Keywords. The author(s) should pick words that accurately describe
%% the work being presented. Separate the keywords with commas.

\received{20 February 2007}
\received[revised]{12 March 2009}
\received[accepted]{5 June 2009}

%%
%% This command processes the author and affiliation and title
%% information and builds the first part of the formatted document.
\maketitle

\section{Introduction}
\label{sec:introduction}

People who are blind or have low vision (BLV) engage in significantly lower levels of physical activity (PA) than their sighted counterparts, a disparity that has been well-documented across multiple population health studies. The Royal National Institute of Blind People (RNIB) reports that 53\% of BLV adults identify as physically inactive, compared to 27\% of the general population \cite{royal_national_institute_of_blind_people_see_2021}. This pronounced gap has far-reaching consequences. Sedentary behavior among BLV individuals has been linked to elevated risks of obesity, cardiovascular disease, and mental health challenges, ultimately leading to a reduced quality of life and increased healthcare burden \cite{lavie_sedentary_2019}. At the root of this problem is a broader structural deficiency: most exercise opportunities, environments, and tools are not designed with non-visual interaction in mind. Traditional fitness modalities overwhelmingly rely on visual feedback for orientation, pacing, and technique correction, rendering them largely inaccessible or ineffective for BLV users \cite{guerreiro_design_2023}. 

Over the past two decades, there has been a growing interest in developing accessible exercise interventions for BLV individuals \cite{mouatt_use_2020}. Several audio-based exergames \cite{exergaming} have emerged, utilizing directional sound cues to enhance interaction and engagement during PA \cite{nunes_echoes_2024}. In a similar vein, researchers have explored the integration of haptic feedback into interactive systems to improve spatial awareness and reinforce action-response loops for BLV users \cite{abtahi_beyond_2019, cassidy_haptic_2013}. Additionally, technologies like virtual reality (VR) and augmented reality (AR) have been studied for their potential to support these systems by providing immersive, multisensory experiences. These technologies can help BLV individuals navigate and interact with their environment, offering auditory and haptic channels to better understand spatial layouts during exercises and mobility training \cite{espinosa-castaneda_virtual_2021, basori_hapar_2020}. Meanwhile, in the domain of PA, adaptive sports such as blind table tennis \cite{morelli_vi-tennis_2010, kamath_playing_2024}, soccer \cite{wood_testing_2021}, and badminton \cite{kim_sonic-badminton_2016} have demonstrated the social and motivational value of sport for BLV individuals. However, these sports often require sighted facilitators, physical infrastructure, and controlled environments, which limit their scalability and autonomous use.
Despite these advances, significant gaps remain. Current VR research for BLV individuals has prioritized assistive navigation rather than physical engagement or exertion \cite{amemiya_navigation_2009, boboc_omnidirectional_2013, siu_virtual_2020, adhikari_lean_2021, lecuyer_homere_2003, todd_virtunav_2014}. Existing exercise applications, while sometimes engaging, often lack spatial adaptability — leaving users unable to intuitively orient themselves within the virtual environment \cite{rector_exploring_2015}. As a result, movement patterns are constrained, gameplay feels disjointed, and the intensity of engagement remains low. More critically, there is an sparsity of VR exergames that explicitly aim to support moderate
to vigorous physical activity (MVPA) \cite{faric_virtual_2021, faric_potential_2022}, a key threshold for health-promoting exercise. A lack of attention to exercise intensity and sustained physical effort can become problematic, given that many BLV individuals already face compounding barriers to fitness and mobility.

In this study, we present our VR boxing exergame \textbf{\textit{PunchPulse}}, designed specifically for BLV individuals. Boxing was selected not only for its rhythmic and full-body movement patterns, but also for its potential to support outcomes across physical (e.g., strength, balance, cardiovascular endurance) \cite{coughenour_changes_2021, sauchelli_virtual_2022, stewart_actual_2022}, cognitive (e.g., attention, spatial reasoning) \cite{costa_virtual_2019, prosperini_exergames_2021}, and psychological (e.g., confidence, self-efficacy, motivation) \cite{coughenour_changes_2021, senn_analysis_2024} domains. We clarify that \textit{\textbf{PunchPulse}} is not the first system of its kind but contributes a distinct focus on sustained MVPA through structured, scaffolded intensity and autonomous interaction within VR exergames. While prior work, including \citet{furtado_designing_2025}, explores rhythm and engagement, \textbf{\textit{PunchPulse}} targets physiological exertion through heart-rate-informed design through round-based increased PA progression. We describe \textbf{\textit{PunchPulse}} as supporting in-game autonomy: BLV players can independently navigate, progress, and engage in combat using built-in game aids. However, the initial setup, including headset positioning and boundary calibration, still requires sighted assistance due to current hardware limitations.

Leveraging the immersive and spatially adaptive nature of VR, our goal is to provide BLV users with a physically engaging, independently usable, and multisensory fitness experience. \textbf{\textit{PunchPulse}} was co-designed over seven months in collaboration with accessibility researchers, including BLV individuals. Our iterative design process focused on eliminating visual dependency through the integration of spatial audio, haptic cues, and intuitive and known interaction paradigms, while also ensuring the system could facilitate PA at intensities aligned with public health guidelines \cite{tanaka_age-predicted_2001}.
% \subsection{Research Questions}
Specifically, this study aims to address the following research questions:

\begin{itemize}
    \item [RQ1.] In what ways can we design a VR boxing exergame that is accessible and engaging for BLV individuals?
    \item [RQ2.] To what extent does our game \textbf{\textit{PunchPulse}}, enable BLV participants to reach and sustain MVPA levels across calibrated intensity stages?
    \item [RQ3.] How do BLV participants perceive the accessibility, exertion, and spatial orientation of \textbf{\textit{PunchPulse}} sought with gameplay intensity increases?
    \item [RQ4.] How do individual factors such as prior physical activity level and VR experience relate to performance outcomes and physiological effort during gameplay? 
\end{itemize}

This work contributes to the growing field of accessible exergaming by addressing the intersection of immersion, exertion, and inclusivity. By grounding our design in the lived experiences of BLV individuals and evaluating its effectiveness through both qualitative and quantitative methods, we aim to bridge a critical gap in adaptive fitness technology. Ultimately, this study demonstrates the potential of VR as a tool for supporting meaningful, effective, enjoyable, and independent PA among BLV users — laying the foundation for broader accessibility in digital health and exercise interventions.

This paper is structured as follows. Section \ref{sec:related_work} reviews prior literature on PA monitoring in VR, accessibility considerations for BLV individuals, and the physical, cognitive, and psychological benefits of boxing as a modality for exertion. Section \ref{sec:design_procedures_and_goals} details the iterative co-design process and outlines the design goals that informed development. Section~\ref{sec:system_implementation} introduces the system implementation, describing the technical architecture, accessibility-centered design features, and the user interaction flow. Section \ref{sec:evaluation} presents the evaluation methods, beginning with the pilot-test and followed by the structured user study, including participant demographics, apparatus, study procedures, and data collection strategies. Section~\ref{sec:findings-pilottest} reports the findings from the pilot-test which informed our design phase and the second iteration of \textbf{\textit{PunchPulse}} development. Section~\ref{sec:findings-user_test} enlists the findings from our user study (both quantitative and qualitative analyses), including observed gameplay behavior and physiological outcomes. Section~\ref{sec:discussion} interprets these results, drawing broader implications for accessible fitness technologies. Finally, Section~\ref{sec:conclusion} concludes the paper by summarizing the study’s contributions and outlining potential future directions.
\section{Related Work}
\label{sec:related_work}

We begin by exploring prior work in crafting exercise-facilitating VR experiences, as well as accessible and enjoyable VR systems across multiple disability communities, paying special attention to advancements within the BLV community. We also discuss how boxing in particular has been explored as a form of PA to enhance multiple facets of one’s well-being.

\subsection{PA Monitoring in VR Exergames}
\label{subsec:pa_monitoring}

% Use \citet{} command when citing authors within the text: \citet{smith2023} showed that...
% Use \cite{} command when citing work: xx was shown to be true \cite{smith2023}.

PA monitoring within VR exergames represents a growing area of research that combines immersive technology with fitness tracking capabilities~\cite{li_sensor_2024, stewart_actual_2022, naugle_exploring_2024}. Prior work has explored the potential of VR to revolutionize exercise routines, enhance motivation, and provide immersive fitness experiences~\cite{siani_impact_2021, mouatt_use_2020, zhou_using_2020, costa_virtual_2019}. Recent studies have also demonstrated that VR exergames can elicit varying levels of PA intensity~\cite{evans_physical_2021, roglin_exercising_2023}. For instance, \citet{naugle_exploring_2024} found that active VR games can significantly increase heart rate from baseline during gameplay, with some games reaching moderate intensity levels comparable to traditional exercise. Similarly, \citet{evans_physical_2021} observed that certain VR games could achieve moderate intensity via percentage of heart rate reserve (\%HRR), with games requiring whole-body movement showing greater moderate-to-vigorous PA (MVPA) compared to those primarily using arm movements.

A particularly interesting aspect of VR exergaming is its potential to alter perceptions of exertion during PA~\cite{costa_virtual_2019, li_sensor_2024}. \citet{stewart_actual_2022} discovered that participants consistently reported lower ratings of perceived exertion (RPE) compared to their actual exertion levels during VR gameplay. This dissociation between perceived and actual exertion suggests that VR may reduce awareness of physical strain, potentially making exercise feel less strenuous and more enjoyable~\cite{prosperini_exergames_2021, borg_psychophysical_1982}.

Beyond physiological measures, research has also examined the psychological benefits of VR exergames~\cite{siani_impact_2021, halbig_systematic_2021}. Studies have found that VR exercise can improve enjoyment, positive affect, and engagement compared to traditional exercise modalities~\cite{sousa_active_2022, sauchelli_virtual_2022}. Furthermore, \citet{sousa_active_2022} reported that active VR not only elicited MVPA but also improved cognitive performance in sedentary college students. The integration of sensor technologies with VR has further enhanced PA monitoring capabilities~\cite{li_sensor_2024, halbig_systematic_2021}. Moreover, prior work has explored sensor fusion-based VR for physical training, demonstrating how the combination of multiple sensors can provide more comprehensive and accurate assessment of movement patterns, exertion levels, and exercise quality~\cite{li_sensor_2024}.

\subsection{Accessibility for BLV People in VR}
\label{subsec:accessibility_BLV_VR}

While VR has presented promising support for facilitating PA, the accessibility of these systems for BLV individuals has been an ongoing topic of exploration~\cite{kreimeier_two_2020, kamath_playing_2024}. Previous research has explored the use of multimodal input (e.g., haptic cues, auditory feedback) to expand the purely visual nature of VR~\cite{kamath_playing_2024, siu_virtual_2020, morelli_vi-tennis_2010}. Recent advancements have focused on making VR more inclusive for BLV individuals through innovative approaches to nonverbal communication and navigation~\cite{kamath_playing_2024, de_silva_understanding_2023}, including the use of audio and haptics to represent nonverbal behavior such as eye contact, head shaking, and head nodding in social VR environments~\cite{dang_opportunities_2023, picarra_creating_2023}.

Beyond communication, navigation in virtual environments presents another significant challenge for BLV users~\cite{guerreiro_virtual_2020, lanca_speed--light_2024}. A framework based on physical sighted guidance has been developed to enable guides to support BLV users with navigation and visual interpretation in social VR~\cite{lo_valvo_navigation_2021, guerreiro_virtual_2020}. This approach allows users to virtually hold onto their guide and move with them while receiving environmental descriptions. Researchers have also explored the development of accessible recreational activities in VR, such as table tennis, specifically designed with and for BLV individuals~\cite{kamath_playing_2024, morelli_vi-tennis_2010, cheiran_inclusive_2011}. These efforts demonstrate the potential to create playful and engaging VR experiences that are fully accessible, expanding opportunities for BLV people to participate in physical activities within virtual environments~\cite{morelli_vi-bowling_2010, bouri_tai_2020}.

\subsection{Boxing as a Means and Method}
\label{subsec:why_boxing}

Boxing has demonstrated a multitude of benefits across various facets of well-being, making it an effective method for promoting holistic health and personal development~\cite{costa_virtual_2019, prosperini_exergames_2021}. As a form of PA, boxing aligns with contemporary approaches to physical health, viewing health holistically through purposeful, full-body exercises~\cite{naugle_exploring_2024, evans_physical_2021}. The sport’s comprehensive nature addresses multiple facets of wellness simultaneously, creating opportunities for both physical engagement and cognitive enhancement~\cite{sousa_active_2022, asadzadeh_effectiveness_2021}.

\subsubsection{Physical Domain}
\label{subsubsec:whyboxing_physical}

Boxing offers significant physical benefits; as a high-intensity sport, it provides an excellent cardiovascular workout, improving heart health and endurance~\cite{lee_physical_2022, coughenour_changes_2021, qin_physical_2020}. The sport’s dynamic nature engages multiple muscle groups, enhancing strength, coordination, and agility~\cite{stewart_actual_2022, srivastav_impact_2021}. For individuals with certain health conditions such as Parkinson’s disease, therapy-based boxing programs have shown particular promise in improving physical functioning and can positively impact the disease course through repetitive, quick motions~\cite{prosperini_exergames_2021, gokeler_immersive_2016}. The combination of aerobic and anaerobic elements in boxing creates a comprehensive physical training experience that addresses multiple fitness components simultaneously~\cite{nathan_impact_2021, evans_physical_2021}. Additionally, studies examining exergames based on boxing movements have demonstrated significant increases in energy expenditure compared to sedentary activities, suggesting potential applications for fitness promotion~\cite{sousa_active_2022, sauchelli_virtual_2022}.

Recent work by \citet{furtado_designing_2025} also leverages boxing in a VR context for BLV users. Their system shares mechanical similarities with ours in terms of target population, boxing as a modality, and use of the Meta Quest 2 headset and in-game aids (employing audio cues for orientation). However, important distinctions exist. \citet{furtado_designing_2025} primarily focused on usability, learnability, and rhythmic engagement, evaluating how users adapt to and interpret cueing schemes and interaction paradigms. In contrast, our focus is on sustained physiological exertion, specifically achieving and maintaining MVPA levels aligned with public health guidelines \cite{tanaka_age-predicted_2001}. Furthermore, while both systems operate on the Meta Quest 2 platform, we acknowledge that the shared hardware imposes significant accessibility constraints, particularly around initial setup and sensory feedback. The Quest 2 lacks system-wide screen reader support or tactile calibration tools, limiting the extent of autonomous onboarding possible for totally blind users. These constraints shaped our technical decisions and underline the need for future VR hardware to better support accessibility by design. We developed our system completely independent and without any collaboration or feedback from the authors of \citet{furtado_designing_2025}.

\subsubsection{Cognitive Domain}
\label{subsubsec:whyboxing_cognitive}

The cognitive benefits of boxing extend beyond physical fitness~\cite{prosperini_exergames_2021, li_sensor_2024}. The sport demands quick decision-making, strategic thinking, and heightened focus, which contribute to improved cognitive function~\cite{salmhofer_design_2023, choa_shadow_2012}. In therapy-based programs for patients with neurological conditions, boxing has been observed to benefit cognitive functioning alongside physical improvements~\cite{prosperini_exergames_2021, asadzadeh_effectiveness_2021}. The complex movements and rapid responses required in boxing may help maintain and enhance cognitive abilities, particularly in areas such as reaction time and spatial awareness~\cite{senn_analysis_2024, ali_virtual_2017}. The dual-task nature of boxing -- requiring simultaneous attention to offensive and defensive maneuvers -- creates cognitive challenges that may transfer to everyday functioning~\cite{mouatt_use_2020, greinacher_impact_2020}. Research on exergames incorporating boxing elements has shown improvements in cognitive performance following acute bouts of exercise, suggesting potential cognitive enhancement effects~\cite{sousa_active_2022, costa_virtual_2019}.

\subsubsection{Psychological Domain}
\label{subsubsec:whyboxing_psychological}

Boxing has also shown remarkable potential in enhancing psychological well-being~\cite{prosperini_exergames_2021, siani_impact_2021, asadzadeh_effectiveness_2021, srivastav_impact_2021}. The sport can boost self-confidence, reduce stress, and provide an outlet for emotional expression~\cite{coughenour_changes_2021, qin_physical_2020}. Moreover, the structured nature of boxing training, with its emphasis on discipline and incremental skill development, provides a framework that can foster psychological resilience and self-efficacy~\cite{stewart_actual_2022, sauchelli_virtual_2022}. These qualities have been noted to be particularly valuable for individuals facing various life challenges, offering transferable skills that extend beyond the sporting context~\cite{senn_analysis_2024, choa_shadow_2012}.

\section{Design Procedures and Goals}
\label{sec:design_procedures_and_goals}

To answer RQ1 and ensure that the system addresses the nuanced needs of BLV users while supporting meaningful physical exertion, we adopted a participatory, ability-based approach to the design of \textit{\textbf{PunchPulse}}. Grounded in iterative collaboration within a mixed-ability research team, our study process involved multiple phases of co-design and evaluation. Phase One (August-December 2024) involved co-designing the prototype with two BLV team members (T1 and T2). Phase Two featured pilot testing with two external blind individuals to gather early feedback on our prototype for refinement (Section~\ref{subsec:pilot-test}). Phase Three included reflective co-design sessions (January-March 2025) with an additional BLV co-designer for fresh insights (T1, T2, and T6), and Phase Four concluded with a user study involving six participants (Section~\ref{sec:evaluation}). This section details the primary design phases (1 and 3), which guided the core design goals that shaped the system implementation (Section~\ref{sec:system_implementation}) and Figure \ref{fig:co-design} illustrates milestone events, co-designer involvement, and the evolution of gameplay mechanics across time. 

\subsection{Design Procedures}
\label{subsec:design_procedures}

Our design process was grounded in participatory and ability-centric frameworks~\cite{wobbrock_ability-based_2011}, combining co-design and iterative evaluation methods to ensure the system reflected both the lived experiences of BLV individuals and the practical constraints of embodied interaction in VR. We adopted the interdependence-Human Activity Assistive Technology (i-HAAT) model~\cite{lee_interdependence-human_2022} to guide the development of a system that supported user autonomy while recognizing the dynamic interplay between individuals, tasks, and environmental contexts. Alongside this, we applied the Rapid Iterative Test and Evaluation (RITE) workflow~\cite{medlock_17_2005} to quickly identify, respond to, and implement design refinements across sessions.

Our mixed-abilities team, comprised of individuals with diverse visual dis/abilities as illustrated in Table~\ref{tab:vis_acuity}. T1 and T2 who are BLV, were integral to all stages of the design process, offering lived-experience insights into navigation, comfort, and engagement within \textbf{\textit{PunchPulse}}. The remaining team members led implementation, documentation, and system architecture tasks, ensuring accessibility considerations were technically feasible and cohesively integrated.

\subsubsection{Initial Co-Design}
\label{subsubsec:initial_co-design}

From August to December 2024, weekly 60-minute in-person co-design sessions were held in the laboratory at our institution. Using the RITE approach~\cite{medlock_17_2005}, sessions began with a summary of recent updates, new features, and changes based on prior feedback. BLV team members conducted hands-on testing and provided real-time feedback via think-aloud methods facilitated by T1. Each session concluded with reflective discussions to assess progress, address emerging issues, and plan the next steps.

\begin{table*}[ht]
  \centering
  \footnotesize
  \begin{tabular}{p{1.5cm} p{12.5cm}}
    \hline
    \textbf{Team Member ID} & \textbf{Visual Acuity} \\
    \hline
    T1 & Blind with no perception of lights and shapes in the right eye and ability to perceive lights and shapes in the left. \\
    \hline
    T2 & Macular dystrophy resulting in weakened central vision in both eyes with a reliance on peripheral vision. \\
    \hline
    T3 & Not BLV, diagnosed with myopia, wears corrective lenses while testing the game. \\
    \hline
    T4 & Not BLV \\
    \hline
    T5 & Not BLV \\
    \hline
    T6 & Sjogren's Syndrome (SJS), with severe light sensitivity. Slightly weakened central vision in both eyes and can see computer screens and digital displays only for very limited time. \\
    \hline
  \end{tabular}
  \caption{Visual Acuity of Researchers}
  \label{tab:vis_acuity}
\end{table*}

Initial development began with T1 and T2 and focused on building low-complexity prototypes that BLV users could quickly learn and test. The first prototype was a static rhythm puncher, in which enemy attacks followed a fixed beat pattern. While accessible, T1 and T2 expressed that the rhythm-based format quickly felt repetitive and lacked engagement. T1 emphasized the absence of full-body challenge, while T2 noted it made gameplay too predictable. These observations led to a shift toward a dynamic opponent system featuring reactive movement (e.g., sidestepping, ducking and advancing) to better simulate boxing. During this time, several features were explored and later removed. One example was the inclusion of audience background tracks meant to simulate a stadium atmosphere. However, both T1 and T2 found that the background noise interfered with cue detection. Based on their feedback, we eliminated ambient sounds to preserve auditory clarity for essential orientation cues.

The "pull enemy" mechanic, designed to help users manage enemy proximity, was initially suggested by T3. Over time, both BLV co-designers found it counterproductive to our physical activity goals. T2 noted that summoning enemies reduced movement demands and allowed players to remain stationary. The feature was appended in November 2024 and randomized enemy movement was added which allowed the enemy to disappear and reappear at a different location. This phase concluded with a pilot test (Section~\ref{subsec:pilot-test} (December 2024)) involving two external blind participants (PP1 and PP2). This early intervention was crucial for validating our design decisions and guiding the next steps in the design process. Gameplay recordings, verbal feedback, and video observations were collected and analyzed by T3, who presented a synthesis of findings at the next co-design session. This marked a transitional point in the process: the insights from pilot testing helped prioritize refinements related to audio clarity, timing of cues, and user and enemy pacing through gameplay rounds.
Their feedback echoed prior concerns: delayed cue delivery, inconsistent pacing, and unclear round transitions. The pilot also highlighted how performance-based progression led to uneven difficulty, which influenced our next round of design refinements.

\begin{figure}
    \centering
    \includegraphics[width=1\linewidth]{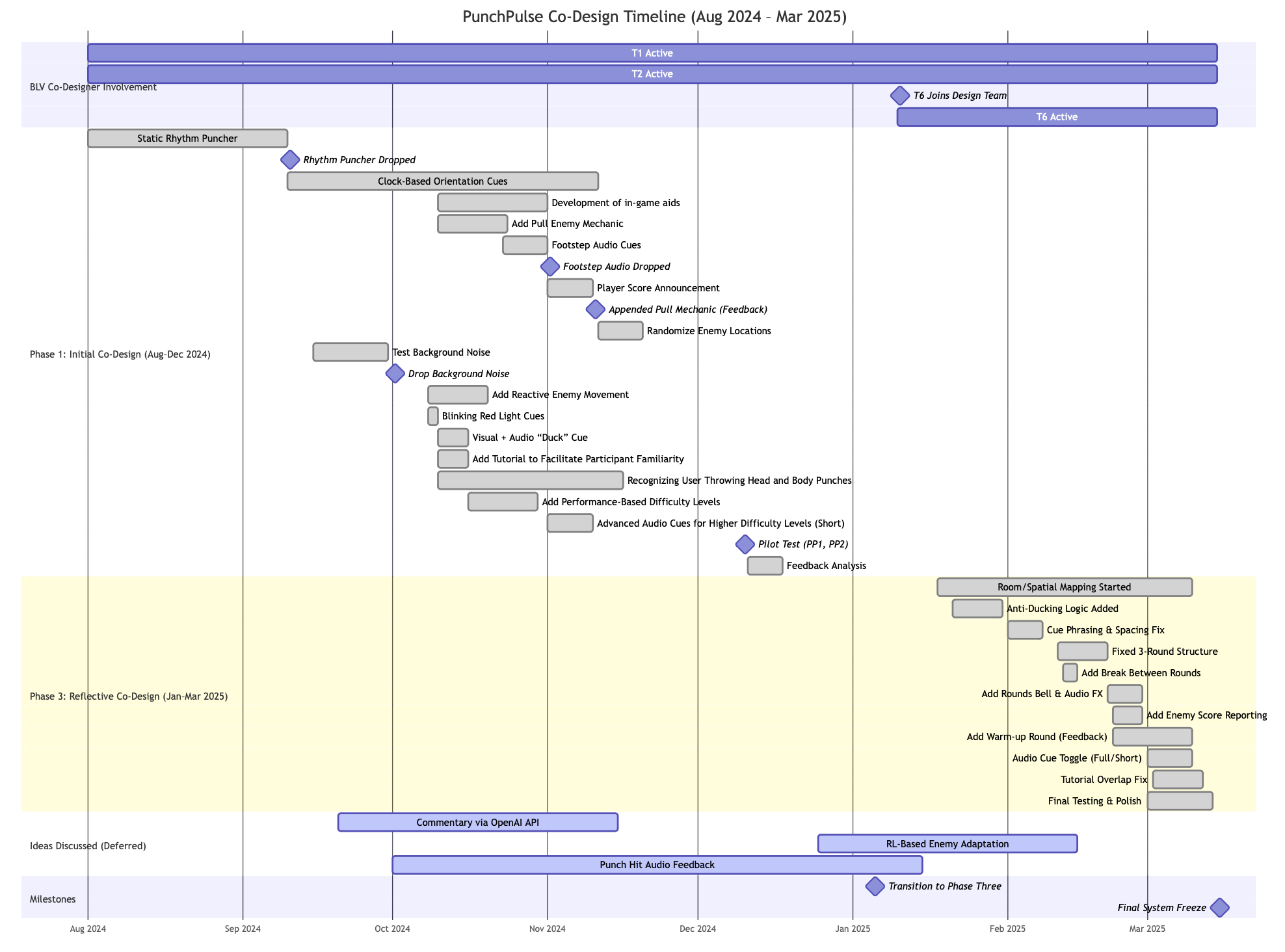}
    \caption{Co-Design Timeline}
    \Description{The PunchPulse co-design timeline spans from August 2024 to March 2025. At the top of the diagram, three horizontal bars indicate co-designer involvement: T1 and T2 are active throughout the entire period, from August to March. T6 joins the design team in January 2025 and remains active until March. Below this, the timeline is divided into Phase 1: Initial Co-Design from August to December 2024, and Phase 3: Reflective Co-Design from January to March 2025. In August, development begins with the Static Rhythm Puncher, which concludes and is marked as dropped in early September. From mid-September to mid-November, Clock-Based Orientation Cues are active. Between early October and early November, the team works on in-game aids, including the Pull Enemy Mechanic, Footstep Audio Cues, and Player Score Announcement. Footstep Audio Cues are dropped on November 1, and the Pull Enemy Mechanic is appended based on feedback and marked with a milestone on November 10. Randomized Enemy Locations are added in mid-November. Test Background Noise runs from mid to late September and is marked as dropped on October 1. Reactive Enemy Movement is added between October 8 and 20. Also in October, Blinking Red Light Cues and Visual plus Audio “Duck” Cues are added, followed by the implementation of a tutorial to facilitate participant familiarity. From October through mid-November, the system adds recognition of user head and body punches. Performance-Based Difficulty Levels are added in mid-to-late October, followed by the implementation of Advanced Audio Cues for Higher Difficulty Levels in early November. On December 10, a Pilot Test with participants PP1 and PP2 is conducted, and feedback analysis follows from December 11 to 18. In January, Room or Spatial Mapping begins on January 18 and continues through early March. Anti-Ducking Logic is added in late January. In February, Cue Phrasing and Spacing Fixes are applied in the first week, followed by the implementation of a Fixed 3-Round Structure, which is in place by mid-February. A Break Between Rounds is added from February 12 to 15. Audio features such as Rounds Bell are added from February 21 to 28. Enemy Score Reporting occurs over the same period, and a Warm-up Round is added from February 22 through early March. An Audio Cue Toggle, allowing full or short cues, is added from March 1 to 10. A Tutorial Overlap Fix occurs from March 2 to 12. Final Testing and Polish is marked from March 1 to 14. In the lower portion of the chart, three long-duration tasks under "Ideas Discussed (Deferred)" are shown. Commentary via OpenAI API spans from September 20 to November 15. RL-Based Enemy Adaptation spans from December 25 to February 15. Punch Hit Audio Feedback spans from October 1 to January 15. Two milestones appear at the bottom of the timeline: Transition to Phase Three on January 5, 2025, and Final System Freeze on March 15, 2025.}
    \label{fig:co-design}
\end{figure}

\subsubsection{Reflective Co-Design}
\label{subsubsec:reflective_co-design}

After the pilot test, our team resumed co-design sessions from January to March 2025, focusing on refining the system based on the feedback received. We invited another blind co-designer, T6, to join our team. T6 had no prior exposure to \textbf{\textit{PunchPulse}}, ensuring the inclusion of a fresh perspective. This addition was instrumental in expanding and reflecting our design and accessibility considerations to a wider spectrum of visual impairments. T6’s feedback was particularly valuable as he approached the system without familiarity, offering critical insights into onboarding, cue clarity, and gameplay strategy. His unique perspective helped the team consider aspects of learnability and development of user strategies while playing the game that may not have been as apparent to recurring testers (T1-T5).

T6 identified issues related to persistent/continuous ducking - players staying low to avoid hits without needing to reorient or physically move. While valid tactically, this undermined the system's physical challenge. T6 raised the concern that this behavior could “flatten” the experience and reduce movement variability. T3 and T4 developed a strategy of preventing users from remaining in a crouched position throughout gameplay by including logic that stops the game if the user is continuously lower than their original height (headset elevation from the marked ground) calculated at the beginning of the game (in tutorial phase). T6 confirmed that this workaround was effective when players stayed low for too long. T6 also experienced difficulty interpreting clock-based orientation cues at high speeds, especially under time pressure. While T1 and T2 had grown accustomed to these over prior months, T6’s challenges led to improvements in cue phrasing, repetition, and temporal spacing. Discussions during this phase also opened new directions for future orientation alternatives (e.g., tactile overlays or adaptive cueing and the inclusion of recognizing multiple styles of throws from the user). 

Meanwhile, the gameplay structure was overhauled. T1 expressed concern that adaptive progression during the pilot led to inconsistent effort levels between users. In response, we implemented a fixed three-round structure, each 10 minutes long with escalating difficulty. T3 helped align these design changes with our technical implementation, and also supported testing of haptic feedback refinements and cue synchronization. By March 2025, the co-design team had finalized the system architecture, and \textbf{\textit{PunchPulse}} was ready for structured evaluation. Design decisions throughout this phase reflect a direct response to co-designer insights and behavioral observations. The participatory process emphasized physical realism, user independence, and structured exertion, ensuring the system was not only accessible but effective in supporting sustained MVPA. 

These Phase Three reflective co-design sessions retained the same 60-minute, lab-based format and continued to follow the RITE structure. Iterative updates were tested in-session by BLV co-designers, whose think-aloud feedback was recorded and collaboratively reviewed using Google Docs. This process ultimately shaped the final system (Section~\ref{sec:system_implementation}) that was evaluated in the user study (Section~\ref{subsec:user-study}).

\subsection{Design Goals}
\label{subsec:design_goals}

% through design process, identify design goals.
% up to 5 design goals here... Each design goal is denoted with DG and use "should" verb. Each goal needs to be concisely supported by evidence/prior work.
Inspired by our iterative co-design procedures, we sought to address key accessibility challenges by implementing multimodal navigation aids, real-time feedback mechanisms, high contrast, and customizable gameplay options. The following design goals (DGs) were developed to foster a responsive, enjoyable, engaging, and functionally realistic environment where players could actively participate in a simulated boxing experience without solely relying on visual cues.

\begin{itemize}
    \item [DG1.] \textit{\textbf{The system should mimic boxing mechanics by creating a novel boxing approach through incorporation of authentic enemy movement, reactive behavior, and physical pacing to promote full-body engagement and facilitate physical exertion.}} Our co-design sessions revealed that BLV people value physical authenticity and variety in opponent behavior to maintain engagement and intensity. Feedback from T1 and T2 led to the development of an enemy logic that mimics real boxing movements such as advancing, sidestepping, and retreating, which increased unpredictability and required the player to stay active and alert. Punch registration was tied to full arm extension rather than button presses to further promote whole-body motion. As gameplay evolved, we appended assistive shortcuts like the 'pull enemy' feature, requiring users to physically step forward to maintain range - intentionally increasing exertion and better aligning with MVPA goals. \textit{This goal draws directly from the design shifts documented in Section~\ref{subsubsec:initial_co-design}, where early concepts such as the static rhythm puncher were replaced through iterative feedback. The removal of the pull mechanic and integration of physical boxing behaviors reflected T1 and T2’s priorities around realism and exertion. These changes are mapped in Figure~\ref{fig:co-design} as major design pivots tied to co-designer contributions.}
    
    \item [DG2.] \textit{\textbf{Spatial orientation and real-time combat awareness should be supported through synchronized auditory and haptic cues, allowing BLV players to locate opponents, respond to attacks, and navigate the environment with minimal cognitive load.}}
    T1 and T2 emphasized the importance of spatial clarity and immediate sensory processing. To address this, we implemented a directional audio system using a clock-face metaphor (e.g., ``Enemy at 3 o’clock''), combined with distance estimates (e.g., ``two steps away''), which allowed players to localize threats. This feedback made spatial mapping intuitive and improved users’ ability to orient themselves and respond effectively under time pressure and PA.\textit{The iterative development of spatial cue systems is described in Section~\ref{subsubsec:initial_co-design} and further refined in Section~\ref{subsubsec:reflective_co-design}, particularly as new challenges were made apparent with T6’s onboarding. Orientation cue updates are explicitly tracked in the project timeline shown in Figure~\ref{fig:co-design}.}
    
    \item [DG3.] \textit{\textbf{The system should enable autonomous play by supporting BLV users in managing timing, defense, and movement, without requiring sighted assistance or external guidance.}} Our participatory design process prioritized autonomy as a core value, with BLV co-designers consistently highlighting the need to engage with the system independently. In response, we replaced visually guided menus with audio-navigable controls, ensured all essential feedback was non-visual, and removed the need for sighted setup during gameplay. The tutorial phase, voiced explanations, and joystick-based UI allowed users to complete onboarding, select levels, and track performance without intervention. The final design enabled participants to initiate and complete rounds while developing their own strategies for defense and movement—supporting fully independent in-game play.
    
    \item [DG4.] \textit{\textbf{Core mechanics such as enemy targeting, punch registration, and round transitions should be designed to maximize immersion and enjoyment for BLV users through continuous sensory feedback and responsive interaction loops.}} The co-design team, particularly T6 during reflective sessions, emphasized the importance of responsive mechanics that reinforced user actions. We observed that participants responded positively to consistent cause-effect feedback, such as enemies audibly reacting to punches and recoiling based on punch intensity. Sonification and verbal cues were also used to differentiate between event types, reinforcing in-game actions and maintaining flow. Round transitions included clear auditory announcements and controlled pauses, allowing users to stay immersed without disorientation. These features collectively created a tight feedback loop that increased player satisfaction and emotional engagement. \textit{This goal builds on the emphasis on sensory clarity and cue timing that evolved in Section~\ref{subsubsec:reflective_co-design}, particularly through T6’s feedback on immersion and control. The integration of consistent punch response and round feedback loops is tied to implementation milestones shown in Figure~\ref{fig:co-design}.}
    
    \item [DG5.] \textit{\textbf{The system should deliver gameplay in sequential time-based stages of increasing intensity, allowing for controlled physical exertion while maintaining consistency across participants for experimental comparison.}} To ensure fair and measurable exertion across users with varying fitness levels, we transitioned from performance-based progression (used in the pilot) to a fixed three-round model with escalating difficulty. Each round lasted ten minutes, separated by brief rest periods, and increased in pace, enemy aggressiveness, and cue sharpness. This time-structured model was informed by co-designer input (e.g., concerns about fatigue and unpredictability) and ensured all participants experienced a consistent intensity increase. Heart rate data visualized in Figure \ref{fig:mvpa} confirmed that this design reliably moved players into MVPA zones, supporting the goal of measurable, scalable physical engagement. \textit{This progression is a direct response to pilot feedback described in Section~\ref{subsec:pilot-test} and was operationalized through reflective co-design in Section~\ref{subsubsec:reflective_co-design}.}
\end{itemize}
\section{System Implementation}
\label{sec:system_implementation}

\subsection{Overview}
\label{subsec:system_overview}

\begin{figure}[ht]
    \centering
    \includegraphics[width=1\linewidth]{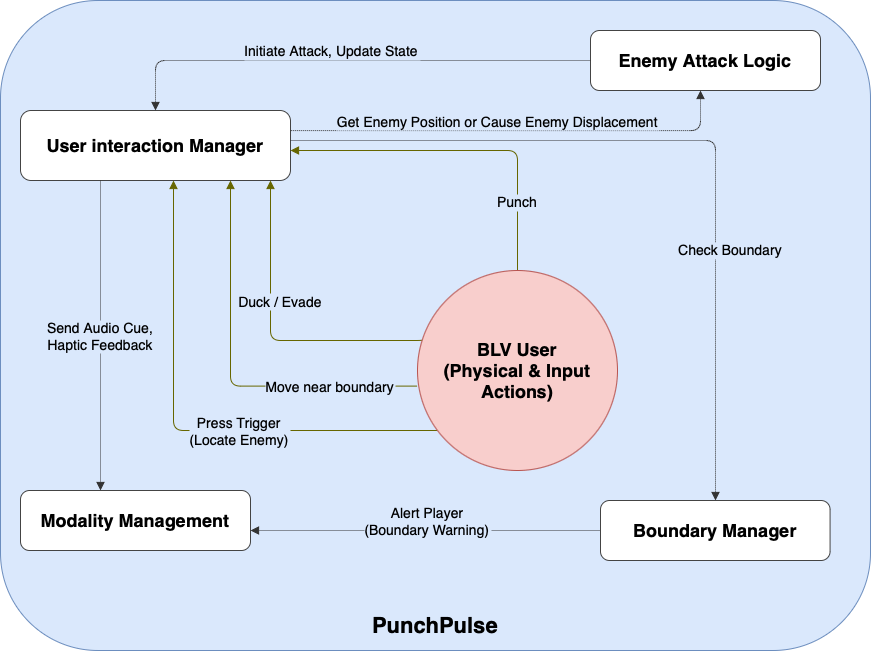}
    \caption{System Architecture and User Workflow}
    \Description{The system architecture diagram shows the interaction between a BLV user and four main system modules within PunchPulse (User Interaction Manager, Modality Management, Enemy Attack Logic, Boundary Manager). At the center is the BLV user, who performs physical actions such as punching, ducking, and moving. In the top left is the User Interaction Manager, which processes these actions and provides feedback through audio and haptic cues. Bottom left is the Modality Management module, which sends these cues to the user and receives updates from the Interaction Manager. On the top right is the Enemy Attack Logic module, responsible for initiating attacks and updating the enemy state. It communicates with both the Interaction Manager and the user. At the bottom right is the Boundary Manager, which checks player positioning and triggers alerts when the user nears the edge of the play area. These modules work together to support real-time interaction and spatial awareness for the BLV user in PunchPulse through non-visual feedback.}
    \label{fig:sys_arch}
\end{figure}

\textbf{\textit{PunchPulse}} is implemented and hosted on Unity and optimized for the Meta Quest 2 platform. The system integrates spatial audio, dynamic haptics, and progressive levels to simulate an immersive and accessible boxing experience. Figure \ref{fig:sys_arch} illustrates a schematic representation of the system architecture and user interaction flow. Further, all refinements made following the pilot-test -- such as round-based game workflow, warm-ups, and audio cue restructuring -- were directly informed by participant feedback and analyzed thematically by T3.

\subsection{Architecture}
\label{subsec:system_architecture}

\subsubsection{Key Accessibility Features and Design Contributions}
\label{subsubsec:design-contributions}

The boxing system centers around embodied interaction, requiring players to punch, duck, and move in response to dynamic in-game stimuli. We implemented several accessibility aids that would help BLV players independently play \textbf{\textit{PunchPulse}} after initial introduction with hardware features. A list of these aids can be found in Table~\ref{tab:controls}. To encourage physical exertion and simulate the demands of real-world boxing, the enemy was programmed with natural boxing movements -- advancing, retreating, and sidestepping within the virtual ring (an introduction as a result of the pilot-test). This enemy behavior were designed to create variability in gameplay, ensuring players could not predict attacks and were therefore required to remain physically engaged and active (\textit{DG1}). Complementing this, punch registration is gesture-based rather than button-activated, requiring users to fully extend their arms toward the enemy, promoting full-body movement and aligning in-game success with physical motion (\textit{DG1}).

\begin{table*}[ht]
\footnotesize
\centering
\caption{Navigational Aids and User Interactions}
\label{tab:controls}
\begin{tabular}{p{3cm} p{3.2cm} p{8.5cm}}
\hline
\textbf{Control} & \textbf{Button/Action} & \textbf{Description} \\
\hline
Locate Enemy & Left Trigger & Plays spatial audio cue to locate the enemy. \\
\hline
Pull Enemy Closer & Right Trigger & Brings the enemy closer for interaction. \\
\hline
Play Score & Left Grip Button & Announces the player’s current score aloud. \\
\hline
Navigate Menu & Left Joystick & Scrolls through menu options, with audio descriptions for each item. \\
\hline
Select Menu Option & Joystick Button (Press) & Selects the currently highlighted menu option. \\
\hline
Open/Close Menu & Left Primary Button & Opens or closes the in-game menu. \\
\hline
Duck/Evade Enemy Attack & Physical Duck & Physically duck to avoid attacks when prompted. \\
\hline
Punch Enemy & Physical Punch & Punch toward the enemy; vibrations confirm successful hits. \\
\hline
\end{tabular}
\end{table*}

To support spatial awareness and orientation without complete reliance on visual feedback, a clock-based directional audio system was implemented. When players press the left trigger, they receive an audio cue indicating the enemy’s relative position in the form of clock-face directions and approximate step distance between the enemy and player. These cues are tightly synchronized with enemy movement and delivered through spatial stereo output, allowing players to turn and face the correct direction without needing to memorize layouts or rely on visual landmarks (\textit{DG2}) as suggested by T1 and T2. To avoid auditory overload, especially when multiple cues are present, the system implements a dynamic layering strategy that staggers competing sounds without masking essential information (\textit{DG2}). This feature involves audio cutting (when prioritized cues are triggered and need to be timely delivered when another cue is in progress) and audio dimming to allow prioritized cues to be expended louder than trivial ones. In addition to auditory orientation, the system also provides tactile feedback to reinforce spatial and combat cues. Distinct haptic patterns are delivered through the controllers when the player lands a hit, gets hit, or approaches the ring's boundaries (suggested by T1). 

\begin{figure}[ht]
    \centering \includegraphics[width=1\linewidth]{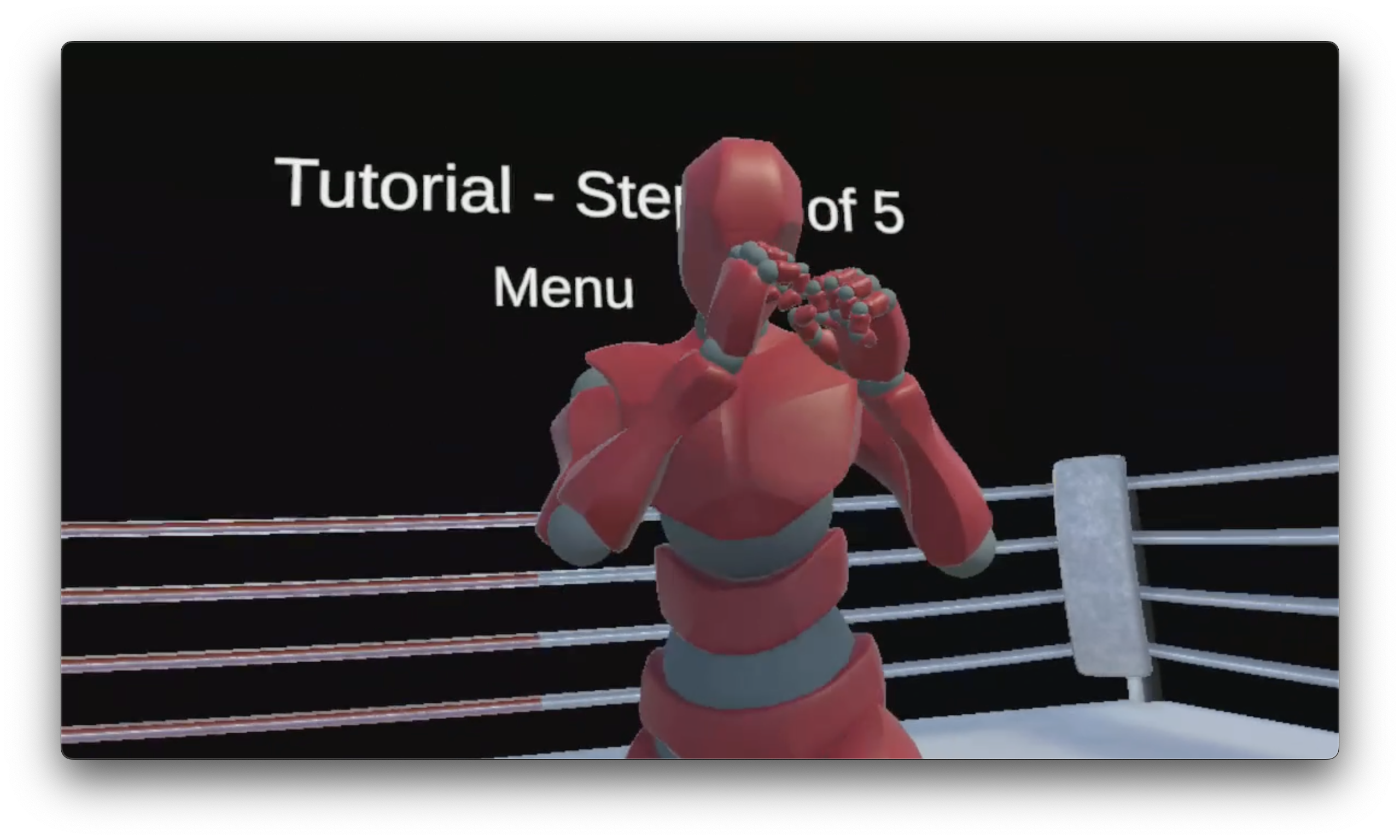}
    \caption{\textbf{\textit{PunchPulse}} Virtual Environment}
    \label{fig:ve_high-contrast}
    \Description{A screenshot from PunchPulse's virtual environment. The current scene features a bright red humanoid opponent (enemy) standing in a boxing stance inside a ring with high-contrast white ropes against a black background. The opponent has distinct body segments and raised fists, making them clearly visible. White bold text at the top reads, "Tutorial - Step 1 of 5 Menu," indicating the tutorial stage. The game's design uses strong color contrast and simple shapes to enhance visibility and focus towards the humanoid.}
\end{figure}

These vibrations differ in intensity and duration based on the nature of the event (\textit{DG2}). For example, a higher velocity punch thrown by a user instantiated higher intensity vibration through the Quest 2 controllers as designed by T3 and T4 based on feedback from T6. These real-time haptic responses not only enhance combat awareness but also help reduce reliance on memory or narration, allowing the player to respond fluidly (via muscle memory developed through game familiarity) during gameplay (\textit{DG4}). We also implemented a high-contrast virtual environment in which low-vision players could easily discern the opponent, in case they did not want to solely rely on audio cues (\textit{DG2}) based on Phase One feedback from T1. A screenshot of the high contrast in our system can be seen in Figure~\ref{fig:ve_high-contrast}.

To enable autonomous play, the system avoids any requirement for sighted guidance during setup or play. All hardware-based controls are introduced by researchers at the start of the session, explaining joystick inputs, combat mechanics, and scoring. We recognize that in order to make the system open-source, we must convert researcher explanations into an audio narration to be played before the tutorial. The in-game menu is navigable via joystick with verbal feedback confirming selections. As a result, users are able to complete tutorials, initiate rounds, and access feedback entirely on their own (\textit{DG3}). The core game loop was carefully designed to support independent timing and reaction; enemy attacks are signaled via predictive and intuitive audio cues, and players are given full control over their defensive responses, enabling them to develop personalized strategies without external support (\textit{DG3}).
To maintain immersion and support long-term engagement, the system leverages a continuous feedback loop between user input and environmental response. Each punch, dodge, or movement triggers an immediate reaction from the enemy or environment -- either through audio response, vibration, or enemy repositioning. For instance, if the player lands a punch, the enemy emits a distinct reaction sound and is physically pushed back (distance of the animation is determined by the vigor with which the player punches the enemy) and bends over, while the controller provides a confirming vibration (\textit{DG4}). Similarly, when the player is hit, the system communicates this through a thud sound effect and a decrement in the player round score, creating a strong link between action and consequence (\textit{DG4}), helping generate incentive for players.
Safety and spatial awareness were critical considerations during development. To prevent users from unintentionally stepping out of bounds, the system monitors player position relative to the defined ring space and emits audio warnings when the user approaches a boundary. The system scales dynamically within the guardian space provided by the Meta Quest setup, adapting the ring dimensions to the safe play area. These features were added after the pilot-test, when testers reported uncertainty about space constraints and occasional overextension during gameplay (\textit{DG2}).

A key innovation in the final implementation was the restructuring of progression mechanics. While the pilot-test version used a performance- or time-triggered progression system, the user study employed a fixed three-round structure, with each round lasting ten minutes and increasing in difficulty. This structure ensured all participants experienced escalating intensity while maintaining experimental control (\textit{DG5}). Between rounds, players were given one-minute breaks to simulate boxing intervals and allow for recovery. The change was introduced to prevent fatigue from accumulating unpredictably across users and to standardize exertion periods for physiological measurement (\textit{DG5}).
Several gameplay features were also modified to enhance exertion demands. The “pull enemy” feature, which had previously allowed players to summon the enemy closer, was disabled in the user study build for Round 2 and 3 to make sure participants engaged in intense physical activity rather than use the feature strategically. This was a decision founded in the observations made in differing participant strategies during our pilot-test. Effectively, this required players to physically walk toward the enemy each time, significantly increasing movement and reducing reaction time (\textit{DG1}). Additionally, audio cues were made shorter and more abrupt to reflect faster-paced enemy behavior, thereby requiring quicker reaction and response from players (\textit{DG5}). These adjustments resulted in a more physically challenging and cognitively engaging experience, based on insights drawn from the pilot-test feedback and analyses.

\section{Evaluation}
\label{sec:evaluation}

\subsection{Pilot-Test}
\label{subsec:pilot-test}

\begin{table*}
\centering
\begin{threeparttable}
\footnotesize
\caption{Pilot-Test Participant Demographics}
\label{tab:participant_demo}
\begin{tabular}{p{20pt} p{20pt} p{30pt} p{60pt} p{50pt} p{50pt} p{120pt}} 
  \hline 
  \textbf{PID} & \textbf{Age} & \textbf{Sex} & \textbf{Job} & \textbf{Used VR Before} & \textbf{VR Knowledge} & \textbf{Visual Acuity} \\ 
  \hline 
  PP1 & 27 & Male & Doctoral Student & Yes & Intermediate & Blind with no perception of lights and shapes in the left eye and bilateral retinal detachment with acuity of 20/2000 in the right eye. \\ 
  \hline
  PP2 & 33 & Male & Unemployed & Yes & Basic & Blind with Leber congenital amaurosis – having ``swiss cheese vision'' and no central perception. \\ 
  \hline 
\end{tabular}
\begin{tablenotes}
\footnotesize
\item[] \hfill \textbf{Note:} ``PP'' indicates pilot-test participant.
\end{tablenotes}
\end{threeparttable}
\end{table*}

To assess the accessibility and physical engagement potential of the VR boxing system prior to full-scale testing, a pilot-test was conducted with two blind participants identified through established community contacts. Participants were screened to meet the U.S. legal definition and NFB's social definition\footnote{https://nfb.org/sites/default/files/images/nfb/publications/fr/fr19/fr05si03.htm} of blindness or low vision, be over the age of 18, and demonstrate an interest in PA. Both individuals had prior exposure to VR technologies -- one with intermediate familiarity through previous accessibility-focused studies, and the other with basic experience. This range of exposure allowed the team to evaluate the system’s usability for different levels of prior VR knowledge (Table~\ref{tab:participant_demo}).

The pilot-test took place in a controlled environment equipped with the Meta Quest 2 headset and controllers with two BLV individuals whose visual acuity has been included in Table~\ref{tab:vis_acuity_participants}. Participants’ gameplay was recorded using multiple modalities: real-time screencast capture from the Meta Quest 2 headset and mobile phone backups. This multimodal capture allowed researchers to triangulate participant behavior, verbal feedback, and physiological responses. Additionally, participants wore their personal Apple Watches during gameplay to monitor heart rate and energy expenditure, enabling analysis of physical intensity during sessions. Gameplay lasted approximately 17 minutes, with a built-in threshold requiring users to reach 20 points in two minutes to automatically advance to higher difficulty levels. If this threshold was not met, difficulty was adjusted manually by the player upon researcher intervention, through the in-game menu. This mechanic enabled the team to observe variations in interaction strategies and physical effort without penalizing users for slower initial adaptation.

Each session began with a brief tutorial on game controls and mechanics, including an explanation of directional audio cues and punch registration. 
The study space used for the pilot-test was larger (approximately 270 sq.ft. out of 440 sq.ft. was used for the study) than the original co-design room, allowing participants to move more freely and reducing the risk of collisions. 
Following gameplay, participants completed the ITC-SOPI survey~\cite{lessiter_cross-media_2001} to assess perceived immersion, engagement, and usability, and participated in a semi-structured interview to elaborate on their experience. However, we chose not to include the SOPI data collection for our user study protocol and do not report the results in Section~\ref{sec:findings-pilottest}, as they were largely repetitive and did not meaningfully enhance our understanding of subjective usability beyond insights already captured through interviews and observational data. Further, this phase revealed a bug where scores were no longer announced past 45 points, requiring researchers to verbally intervene -- an issue that was resolved before user testing. 

\subsection{User Study}
\label{subsec:user-study}

\subsubsection{Participants}
\label{subsubsec:participants}

With approval from our Institutional Review Board (IRB), we recruited six blind and low-vision participants (visual acuity descriptions can be found in Table \ref{tab:vis_acuity_participants_user-test}) for our user study using snowball sampling through community and personal networks. Participants were compensated with a \$50 Amazon gift card for their time. The study was conducted in-person and lasted approximately one hour per participant. Only one participant (P2) had prior exposure to \textbf{\textit{PunchPulse}} during earlier prototype pilot testing.

\begin{table*}[ht]
\centering
\begin{threeparttable}
\footnotesize
\caption{Visual Acuity of User Test Participants}
\label{tab:vis_acuity_participants_user-test}
\begin{tabular}{p{0.8cm} p{16cm}} % adjusted for full-width
  \hline
  \textbf{PID} & \textbf{Visual Acuity} \\
  \hline
  P1 & Blind in both eyes with acuity of 10/200. \\  
  \hline
  P2 & Blind with no perception of lights and shapes in the left eye and bilateral retinal detachment with acuity of 20/2000 in the right eye. \\  
  \hline
  P3 & Blind with no perception of lights and shapes in the left eye and low-vision with an acuity of 10/140 in the right eye. \\ 
  \hline
  P4 & Completely blind with no perception of lights and shapes in the left eye and low-vision in the right eye. \\ 
  \hline
  P5 & Both eyes have medically diagnosed shape, color and light perception; however, when they cover their right eye, the left eye is blank. \\ 
  \hline
  P6 & Totally blind with no perception of lights and shapes. \\ 
  \hline
\end{tabular}
\begin{tablenotes}
  \footnotesize
  \item[] \hfill \textit{Note:} PP1 and P2 are the same person.
\end{tablenotes}
\end{threeparttable}
\end{table*}

To ensure an appropriate and diverse sample, participants were required to meet the following inclusion criteria:
\begin{enumerate}
    \item Must be 18 years of age or older.
    \item Must meet the U.S. legal definition and NFB's social definition of blindness.
\end{enumerate}

Participants self-reported their PA levels and prior VR exposure. PA was categorized into three levels: Highly Active, Somewhat Active, and Never Workout. VR experience ranged from no exposure to VR to prior experience with \textbf{\textit{PunchPulse}}. The participant group consisted of five males and one female, with ages ranging from 27 to 42 years ($\bar{x} = 31.67, \textit{s} = 6.02$) and education levels ranging from high school to university graduates. Table~\ref{tab:participant_demo_exp} provides an overview of participant demographics, PA levels, VR experience, and whether a warm-up phase was necessary based on initial gameplay response.

\subsubsection{Apparatus}
\label{subsubsec:apparatus}
To deliver a safe, immersive, and accessible VR boxing experience for BLV participants, we employed a multi-device apparatus setup that emphasized reliable motion tracking, non-visual feedback, and consistent session execution. The core system was deployed on the Meta Quest 2 standalone headset, selected for its robust 6DoF spatial tracking~\cite{banaszczyk_6dof_2024}, integrated haptics, and stereo audio rendering. Participants used the default Quest handheld controllers, which provided both haptic feedback and accurate capture of upper-body movement for punching and navigation. For physiological monitoring, a single lab-owned Apple Watch was used across all sessions. The device was paired with a dedicated lab iPhone to ensure standardized configuration and eliminate variation introduced by personal devices. The Apple Watch was worn on the participant’s left wrist throughout the warm-up and all gameplay rounds.

Six white canes were placed on the floor in the real world to create a boundary that would be easily discernible to participants when wearing the headset. The virtual environment was projected using the Quest's guardian boundary system which matched the real-world pre-defined white cane boundary. Prior to each session, researchers remapped the guardian boundary to this cleared section of the room, enabling participants to move freely while avoiding collisions or overextension. The study was conducted in the spacious living room (approximately 250 sq.ft.) of one participant’s home, which closely matched the pilot-test room in size and layout. This consistency in environmental setup across participants ensured comparable freedom of movement and spatial interaction during all sessions. Physical movement data was captured using an externally mounted mobile device positioned at a fixed angle. In-headset screencast recordings were also enabled to record first-person perspective, including gameplay, spatial orientation, and system feedback. These recordings were primarily used to support post-session review and qualitative annotation of gameplay dynamics. The overall apparatus ensured participants could interact with the system independently and safely, while preserving consistency and reliability across all study sessions.

\subsubsection{Study Procedure}
\label{subsubsec:procedure}
Each session followed a standardized within-subjects protocol designed to evaluate how participants responded to increasing gameplay intensity in terms of both physical exertion and interaction experience. Upon arrival, participants provided informed consent and were seated for a five-minute resting period to establish individual baseline heart rates. This value was later used to determine personalized exertion zones for MVPA analysis. Once baseline measurements were complete, researchers fitted the Meta Quest 2 headset and controllers, configured the VR guardian boundary, and initiated the in-headset recording. 
The session began with a short, guided tutorial within \textbf{\textit{PunchPulse}}, introducing players to the core mechanics -- enemy tracking, blocking, striking, and cue interpretation. Following the tutorial, participants completed a warm-up round designed to ease them into the interaction tempo. This phase included light physical actions and used the previous pilot-test “easy” difficulty level, which was not scored in the final evaluation. The main study consisted of three 10-minute gameplay rounds of increasing difficulty, separated by brief rest intervals. The Apple Watch workout was manually started by the researchers using the "Kickboxing" mode to ensure consistent labeling for downstream data filtering. If the participants accidentally crossed the white cane boundary during the rounds, the researchers intervened with a sharp "Stop" audio cue in order to prevent injuries. These rounds were presented in a fixed order to maintain consistency across participants and facilitate comparative analysis. At the end of the final round, researchers stopped the screencast and wearable recordings. Participants then completed a post-study System Usability Scale (SUS) questionnaire~\cite{affairs_system_2013} followed by a semi-structured interview. Interview questions focused on perceived exertion, clarity of spatial feedback, independence during play, and overall usability. The complete procedure, including onboarding, gameplay, and feedback collection, took approximately one hour per participant.

\begin{table*}[t]
\centering
\footnotesize
\caption{User Study Participant Demographics}
\label{tab:participant_demo_exp}
\begin{tabular}{p{0.6cm} p{0.8cm} p{0.8cm} p{3.8cm} p{2.2cm} p{2.5cm} p{3.3cm} p{1.2cm}} 
  \hline 
  \textbf{PID} & \textbf{Age} & \textbf{Sex} & \textbf{Highest Level of Education Attained} & \textbf{Prior Exercise Level} & \textbf{Physical Activity} & \textbf{Prior VR Exposure} & \textbf{Warm-up Required} \\ 
  \hline 
  P1 & 28 & Male & Professional Degree & Easy & Somewhat Active & Never Tried VR Before & Yes \\ 
  \hline 
  P2 & 27 & Male & Graduate & Easy to Moderate & Somewhat Active & Tried this Game Before & Yes \\ 
  \hline 
  P3 & 28 & Male & High School & Hard & Highly Active & Tried AR Before & Yes \\ 
  \hline 
  P4 & 36 & Female & Undergraduate & Moderate & Somewhat Active & Never Tried VR Before & No \\ 
  \hline 
  P5 & 29 & Male & High School & Hard & Somewhat Active & Never Tried VR Before & No \\ 
  \hline 
  P6 & 42 & Male & Undergraduate & Moderate & Somewhat Active & Never Tried VR Before & Yes \\ 
  \hline 
\end{tabular}
\end{table*}

\subsubsection{Data Collection and Analysis}
\label{subsubsec:data_collection-analysis}

Data collection targeted physiological, behavioral, and subjective measures to evaluate participant exertion, usability, and spatial orientation across gameplay. Physiological data were gathered via the Apple Watch, which continuously recorded heart rate and active energy expenditure. Minute-level heart rate data were exported using the Health Report app on the paired iPhone. For each participant, we calculated average and peak heart rates per round, as well as total calories burned. 

We chose MVPA as the basis for quantitative evaluation because it provides a well-established benchmark for meaningful physical exertion with known health implications \cite{ramulu_real-world_2012}. Unlike step counts or general movement tracking, MVPA reflects sustained intensity levels that align with PA guidelines and cardiovascular outcomes. This made it a relevant and interpretable measure for assessing whether the system could go beyond encouraging movement and actually support exertion levels associated with health benefits. Given that many BLV-accessible activities emphasize light-intensity exercise~\cite{haegele_health-related_2017}, using MVPA as a metric allowed us to evaluate whether our system could bridge that gap and support more vigorous engagement in a structured, measurable way. 

Time spent in MVPA zones was determined using the Heart Rate Reserve (HRR) method~\cite{tanaka_age-predicted_2001}, with resting heart rate measured pre-session and maximum heart rate estimated as $HR\textsubscript{max} = 208 - 0.7 \times \text{age}$. Moderate and vigorous intensity thresholds were set at $\geq 60\%$ and $\geq 80\%$ HRR, respectively, enabling personalized MVPA analysis.
System logs captured event-level data, including scoring, hits taken, punches landed, and aid activation counts (Table~\ref{tab:player_scores}), supporting triangulation of observed behaviors. Subjective data were collected through the SUS survey and semi-structured interviews. The SUS provided a standardized usability measure, while interviews explored perceptions of exertion, immersion, enjoyment, and spatial cue clarity. Interviews were audio-recorded, transcribed, and thematically analyzed using interpretative phenomenological analysis~\cite{Smith2006IPA}. Two researchers independently conducted open coding, then collaboratively refined codes into thematically related clusters and superordinate themes, grounded in participant narratives and aligned with research questions. Inter-rater reliability was assessed using Cohen's kappa ($\kappa$)~\cite{cohen_coefficient_1960}, achieving a value of 0.828, indicating strong agreement. To minimize bias, triangulation with gameplay behavior and reflexive journaling supported methodological rigor.
Participant behavior was further analyzed through screencast footage, external recordings, and detailed observation notes. Data were segmented into tutorial, warm-up, and gameplay rounds to assess adaptation to mechanics, interaction with spatial cues, and physical response to increasing difficulty.
These multimodal data sources enabled a comprehensive evaluation of the system’s effectiveness in promoting exertion, maintaining engagement, and supporting BLV players’ interaction.
\section{Pilot-Test Findings - Answering RQ1 and RQ3}
\label{sec:findings-pilottest}
This section addresses RQ3 by exploring how BLV participants perceived the initial implementation of \textbf{\textit{PunchPulse}} during our pilot-test, with a particular focus on their experiences navigating its accessibility and spatial orientation features. Since this testing informed our second phase of co-design and development (Section~\ref{subsubsec:reflective_co-design}), we use its findings to answer RQ1 as well. However, because the pilot-test was limited to identifying accessibility issues in the first iteration of \textbf{\textit{PunchPulse}}, we do not address the remaining research questions here. These are instead covered in our user testing findings in Section~\ref{sec:findings-user_test}.

As part of the broader study, our pilot-test explored the usability, accessibility, and physical engagement potential for the first iteration of \textbf{\textit{PunchPulse}}. These findings were analyzed and incorporated into the second phase of our co-design sessions (Section~\ref{subsubsec:reflective_co-design}).
In terms of PA and gameplay strategy, \textbf{\textit{PunchPulse}} initially demonstrated its capacity to accommodate diverse playstyles. PP1 engaged at a high intensity, incorporating frequent ducking and aiming for high scores, which led to sustained effort and cardiovascular engagement. In contrast, PP2 employed a more measured strategy, leveraging the enemy-pull feature to control proximity and reduce movement while maintaining interaction. Quantitative metrics from Apple Watch data supported these observations, showing significant variance in energy expenditure and heart rate patterns between the two participants. This variation illustrates the game's adaptability to different physical ability levels and motivational drivers.
Regarding accessibility and immersive engagement, both participants underscored the centrality of multimodal feedback, particularly spatialized audio cues and haptic signals, for real-time orientation and gameplay execution. Clock-based audio cues enabled effective spatial tracking, although overlaps with concurrent audio (e.g., crowd noise or instructions) occasionally compromised intelligibility. Participants appreciated the game’s responsiveness and realism, though feedback suggested improvements such as periodic score announcements and round-based structuring to enhance pacing and immersion. These comments align with the game’s goal of supporting autonomous, motivating exercise experiences. These were incorporated into \textbf{\textit{PunchPulse}} for our user test. 
We also conducted a thematic analysis for participant responses to open-ended questions during the pilot-test which distilled six key themes (refer to Table~\ref{tab:pilot-thematic_analysis-table}): coordination within the virtual space, cognitive-physical balancing, usability, spatial awareness through multimodality, interaction with core mechanics, and immersive motivational factors. PP1 emphasized the importance of aligning virtual and physical boundaries to feel safe during movement (which translated into placing white canes around the room boundary during the user test), suggesting a desire for one-to-one spatial mapping or clearer boundary alerts. Meanwhile, PP2 highlighted the intuitive nature of the accessible menu and text-to-speech feedback, enabling BLV users to navigate configuration settings independently. Both participants identified the gameplay as cognitively engaging and physically beneficial, affirming the potential of \textbf{\textit{PunchPulse}}.
\section{User Test Findings - Answering RQ2, RQ3 and RQ4}
\label{sec:findings-user_test}
RQ1 was primarily addressed during the design and development stages of \textbf{\textit{PunchPulse}}, as discussed in Section~\ref{sec:design_procedures_and_goals}. This included formative research and initial design decisions informed by BLV user needs. As a result, RQ1 is not revisited in the user test findings presented further.

In this section, we report findings from our mixed-methods evaluation of \textbf{\textit{PunchPulse}}, focusing on its effectiveness in promoting physical exertion, supporting spatial orientation, and enabling non-visual interaction for BLV participants. Our user test analysis draws from three primary data sources: physiological metrics captured via wearable sensors, behavioral data from gameplay recordings and researcher observations, and post-session feedback collected through usability surveys and interviews. Together, these user test results illustrate how participants engaged with the system across physical, cognitive, and perceptual dimensions, and reveal both the strengths and limitations of our current design. 

% We have also linked these findings to our posed research questions, with each subsubsection of user test findings, explicitly addressing how the data informs RQ2, RQ3, and RQ4 respectively.

\subsection{Physical Activity Outcomes with \textbf{\textit{PunchPulse}} (RQ2)}
\label{subsection:findings-rq2_mvpa}
\begin{figure}[ht]
    \centering
    \includegraphics[width=1\linewidth]{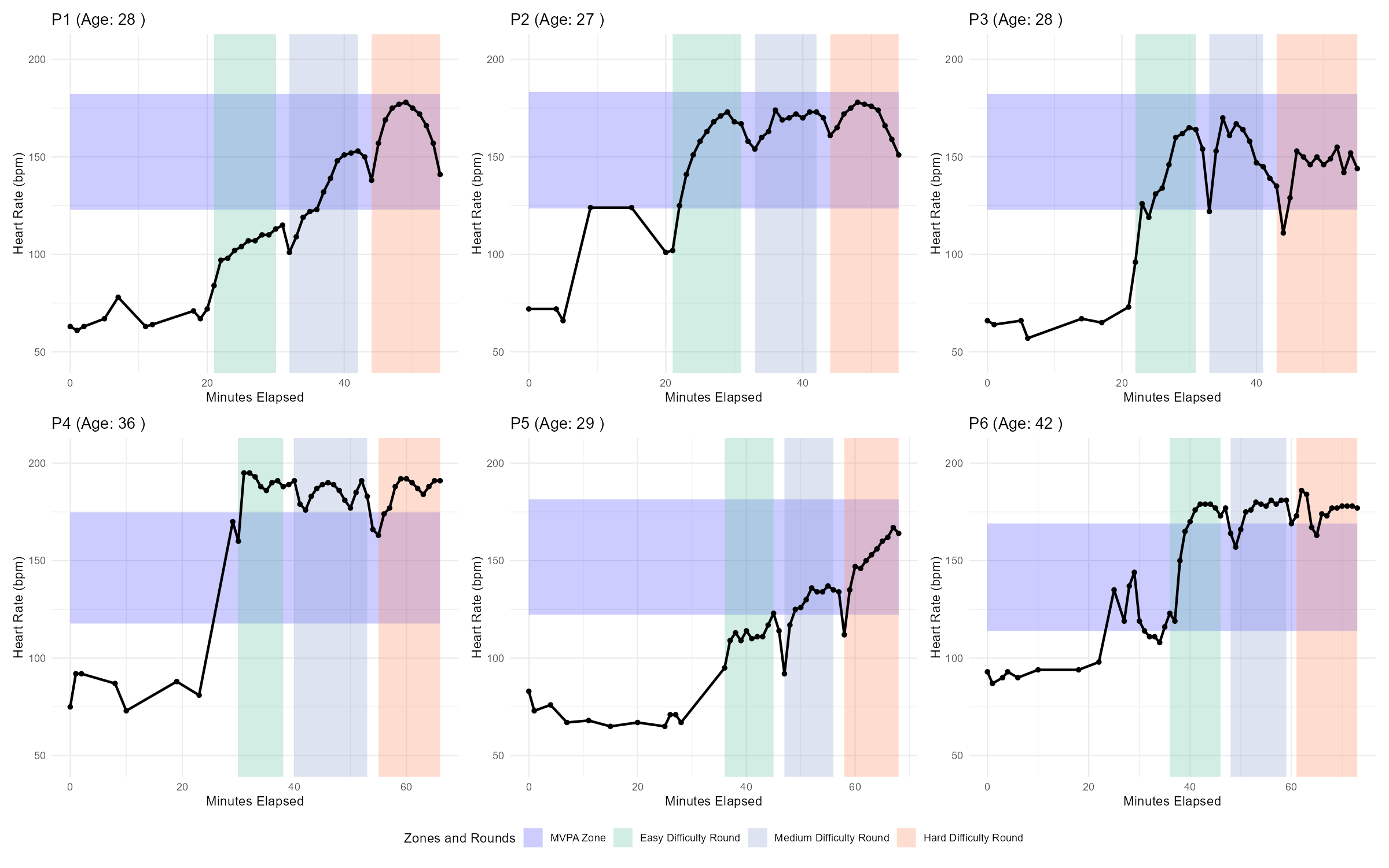}
    \caption{Heart Rate Variability per Participant}
    \label{fig:mvpa}
    \Description{The figure consists of six individual line plots arranged in a grid layout, each representing one participant (P1 to P6). The x-axis of each plot shows time in minutes across the tutorial, warm-up, and three gameplay rounds. The y-axis indicates heart rate in beats per minute. A solid black line, connecting individual heart rate black plot points, represents the participant’s heart rate over time. Behind the line, a horizontal lilac shaded band spans a portion of the y-axis — this band marks the personalized MVPA zone for that participant, calculated using their heart rate reserve. Three vertical spreads along the y-axis indicate the time for which each round spanned per participant, represented by green (easy round), blue (medium round), and orange (hard round) shading. In most plots, the heart rate line begins below the shaded band during the tutorial and warm-up and gradually rises into or above the MVPA zone during gameplay. For P1, P2, and P3, the black line enters the MVPA zone in the easy or medium round and remains within it. P5 enters the zone only briefly during the final round. P4 and P6 display continuous elevation: their heart rate lines rise above the MVPA band early in gameplay and stay above it, showing sustained high exertion. This visualization demonstrates that all participants reached MVPA, and two (P4 and P6) sustained levels well beyond it.}
\end{figure}

Our study of RQ2 was driven by an evaluation of whether participants reached and sustained MVPA levels during gameplay. Using heart rate data mapped to exertion zones, it assesses the effectiveness of \textbf{\textit{PunchPulse}}’s round-based intensity model in promoting and maintaining exertion.
Participants' cardiovascular response over the course of the session revealed clear engagement with MVPA zones, with all participants reaching their personalized exertion thresholds and five of six maintaining them for sustained periods across rounds. The gameplay structure anchored by full-body movement, spatial tracking, and continuous engagement, resulted in progressively increasing heart rates that aligned closely with \textbf{\textit{PunchPulse}}'s intended difficulty ramp. Figure~\ref{fig:mvpa} presents a per-participant visualization of heart rate over time, mapped against round durations and each individual's MVPA zone. The length and density of time spent within these shaded zones provide insight not only into whether thresholds were reached, but how effectively they were retained.

Notably, P4 and P6 exceeded their MVPA thresholds shortly after the start of the easy round and remained well above those boundaries for the entirety of gameplay. For these participants, heart rate remained elevated even during round transitions, suggesting not just responsiveness to physical challenge but cardiovascular endurance. Their curves plateaued at high levels, indicating minimal fluctuation in effort once peak intensity was reached. This continuous retention above the MVPA zone suggests that the system was not only effective in eliciting exertion, but also in sustaining it—an essential criterion for health-promoting exercise. Other participants demonstrated a more graduated trajectory into MVPA. P1 and P2 entered the MVPA zone in the latter half of the easy round and maintained it steadily through the medium and hard rounds. Their heart rate profiles showed a healthy acceleration curve, with each subsequent round yielding slightly higher peaks and longer durations within their target zone. This reflects the efficacy of the time-based difficulty model in guiding users of varying fitness levels into higher intensity zones without the need for calibration. P3 exhibited a more variable heart rate pattern, with intermittent peaks into MVPA that coincided with bursts of movement and vocal expressions of effort. These fluctuations are consistent with a high-intensity interval engagement style, where periods of maximal effort are followed by short recoveries. While less sustained, these peaks nonetheless reflect significant exertion and responsiveness to in-game pacing. P5 presented the most gradual pattern, only entering MVPA late in the session during the hard round. Their data suggest a slower adaptation to \textbf{\textit{PunchPulse}}’s physical demands or a higher baseline threshold. However, the eventual crossing into MVPA, even if delayed, demonstrates that the fixed progression model successfully accommodated a wider exertion range across users.

The distribution of effort across participants highlights the balance achieved between structured game pacing and personalized intensity response. The round-based design, which progressively restricted reaction time and increased enemy movement, produced measurable cardiovascular effects even in participants with no prior VR experience or minimal PA engagement. Importantly, the ability of participants like P4 and P6 to exceed their MVPA zones suggests that the system is capable not only of scaling up to highly active individuals but of sustaining that effort across multiple rounds.

\subsection{Interpreting Users' Accessibility, Exertion and Spatial Orientation in \textbf{\textit{PunchPulse}} (RQ3)}
\label{subsection:findings-rq3-accessibility}
We investigated RQ3 by (1) thematically analyzing participants’ perceptions of exertion, accessibility, and orientation within \textbf{\textit{PunchPulse}}; (2) detailing participants’ real-time behaviors, verbalizations, and orientation challenges during gameplay; and (3) assessing their SUS scores. 

\subsubsection{Thematic Analysis}
\label{subsubsec:findings-thematic_analysis}
Our thematic analysis revealed five interrelated themes: Physical Engagement and Immersion, Sensory Presence, Usability and Interaction Fluidity, Gamification, and Exercise Motivation. The final themes, alongside all cleaned participant quotes, are presented in the Table~\ref{tab:thematic_coding}. These themes were not predefined, but rather emerged from repeated patterns in participant expression—reflecting the lived, affective, and strategic dimensions of interacting with \textbf{\textit{PunchPulse}}.
The theme of Physical Engagement and Immersion captures the way exertion and embodiment became core to the experience, not just as effort, but as immersion. Participants framed their physicality as part of the game’s logic, not its burden. As P2 remarked, ``Sometimes I would swing and it wouldn't hit, but I could feel the distance'', suggesting that success was less about visual precision and more about embodied alignment. The sense of being ``in the game'' stemmed from this sustained motor-sensory loop, where feedback was felt, not seen. Sensory Presence emerged as a distinct dimension, separate from immersion in gameplay. Here, participants described the way audio and haptics situated them in space, helping them form a map of the environment through movement and orientation. While this presence sometimes faltered — e.g., missed cues or feedback delays—it provided a substrate for participants to negotiate space independently. These responses suggest that the system allowed BLV users to reconstruct spatial awareness not by reducing complexity, but by reinforcing alternative modalities. Usability and Interaction Fluidity reflected how well participants could maintain momentum and understanding within the system’s loop of feedback, response, and adjustment. This theme revealed a tension: participants appreciated the fluidity of the experience but also flagged moments of breakdown. As P2 observed, ``There was a few inconsistencies where he just wouldn't come at me'', indicating that perceived usability was as much about system responsiveness as about control layout. These micro-breakdowns were especially salient under high intensity, where participants expected cues to be rapid and reliable.
The theme of gamification encapsulated how scoring, pacing, and rule logic shaped motivation. Participants responded not just to challenge but to fairness and payoff. For example, P1 noted, ``I wish it (enemy) did not pause as much, that way I can go in one rhythm'', expressing a desire for a more game-like cadence. Here, score wasn't merely a performance measure — it was a signal of progress, tempo, and psychological flow. When the pacing aligned with bodily effort, the game became more than exercise — it became strategy. Finally, Exercise Motivation foregrounded the value participants placed on the experience as a fitness tool. \textbf{\textit{PunchPulse}} wasn't just played — it was seen as more effective and engaging than many participants’ usual routines. This suggests that even without relying on visual stimuli, a well-structured multisensory system can transform exercise from obligation to opportunity.

\subsubsection{Behavior Analysis}
\label{subsubsec:findings-behaviour_analysis}
Many participants began the tutorial with hesitancy around control mapping and button actuation. Some required repeated attempts to activate the directional locator correctly or to deliver punches with sufficient intensity. One participant verbalized their frustration mid-tutorial: ``Huh, what the hell, why isn't this working?'' (P6), reflecting an early barrier in coordinating spatial audio feedback with motor actions. This challenge was echoed by another participant, who expressed confusion with the directional cues: ``Clock direction was the most difficult for me'' (P5). 

In several cases, participants initially misunderstood the spatial meaning of directional audio, punching in entirely the wrong direction (P5 and P6) or flailing their arms in broad arcs rather than focused strikes (P4). These challenges can be better understood through three key dimensions we derived: \textbf{(1) \textit{Orientation Strategy Development}}, \textbf{(2) \textit{Cue Misinterpretation Patterns}}, and \textbf{(3) \textit{Engagement Pacing}}.

Issues with \textbf{\textit{Cue Misinterpretation Patterns}} were most apparent during early gameplay, where directional audio feedback failed to translate clearly into effective movement. Participants punched in incorrect directions or misread the clock-based system entirely. This confusion reflects a usability limitation in \textbf{\textit{PunchPulse}}’s spatial cueing system and its reliance on users’ familiarity with abstract spatial conventions like clock orientation.

However, signs of \textbf{\textit{Orientation Strategy Development}} emerged over time. Participants began to narrate their actions and interpret audio cues through self-guided spatial logic. For example, P6 said, ``Okay 11 o'clock, 2 steps away...8 o'clock so back to front'', while later exclaiming, ``3 o’clock—I am here! 90 degrees'', indicating internalized adjustments to orientation. Several participants improvised hybrid strategies using auditory cues, body positioning, and their own estimation of angles and distance. The warm-up phase further revealed individual baselines of physicality. P6 began with energetic front and cossack squats, showing comfort with dynamic full-body movement. P3 similarly engaged \textbf{\textit{PunchPulse}} as ``a fight'', applying competitive focus early on. P4 took an aggressive approach and frequently vocalized the enemy’s movements: ``Come on you!'', ``He is way back there!'', followed by: ``Don’t you laugh at me'' (P2). Despite disorientation, P4 eventually said, ``I think I figured out how it works''—though their accuracy remained poor—highlighting the perception of improvement even when precision did not match.

Changes in \textbf{\textit{Engagement Pacing}} became particularly evident across round transitions. As difficulty increased, participants verbalized shifts in intensity. P2 noted, ``Oh yeah, it moves a lot faster this round'', while P6 remarked, ``Oh come on bro… you start coming at an angle and I can't track it!'' These statements marked not just an escalation in gameplay mechanics but also rising cognitive and perceptual load. P6, previously high-energy, later requested time checks — ``Can you tell me when I’m four minutes away?... two minutes to go...damn it, can’t move'' — and ultimately admitted: ``Might have over exercised myself''. Such utterances revealed a blend of frustration, humor, competitiveness, and fatigue. P6 exclaimed, ``I am ducked...can’t hit me!'' in a defiant moment of exertion, while P2 remarked positively on game design: ``Ooo I like that, so there is incentive''. Although score feedback affected players differently — some found it motivating, while others, like P5, were ``not very motivated by the score itself'' — most participants found the activity rewarding. P5 later described \textbf{\textit{PunchPulse}} as ``a good form of cardio—fun at the same time''. P4 also appreciated visual elements, noting that ``bright colours [were] helpful for the contrast'', although audio and haptics were the primary navigational tools, particularly for low-vision users. Even among those who struggled with orientation or stamina, the system’s immersive structure and rhythm kept participants engaged. As P2 aptly summarized: ``It’s engaging for those who don’t like to exercise''.

\subsubsection{SUS Scores}
\label{subsubsec:findings-sus}
To assess the perceived usability of the \textbf{\textit{PunchPulse}}, we administered the SUS to participants after their play sessions. The SUS is a standardized, 10-item questionnaire that provides a global measure of system usability on a scale from 0 to 100 \cite{brooke_sus_1995}. 

\begin{table*}[h!]
\centering
\caption{Individual SUS item responses and total scores per participant}
\begin{tabular}{lccccccccccc}
\toprule
\textbf{Participant} & \textbf{Q1} & \textbf{Q2} & \textbf{Q3} & \textbf{Q4} & \textbf{Q5} & \textbf{Q6} & \textbf{Q7} & \textbf{Q8} & \textbf{Q9} & \textbf{Q10} & \textbf{SUS Score} \\
\midrule
P1 & 5 & 1 & 5 & 1 & 3 & 3 & 5 & 1 & 5 & 3 & 85.0 \\
P2 & 4 & 3 & 4 & 2 & 4 & 2 & 4 & 1 & 5 & 1 & 80.0 \\
P3 & 3 & 2 & 4 & 3 & 4 & 2 & 5 & 2 & 4 & 1 & 75.0 \\
P4 & 4 & 2 & 5 & 1 & 5 & 1 & 4 & 1 & 5 & 2 & 90.0 \\
P5 & 5 & 2 & 5 & 3 & 5 & 1 & 5 & 1 & 4 & 3 & 85.0 \\
P6 & 4 & 3 & 2 & 2 & 3 & 4 & 4 & 3 & 3 & 2 & 55.0 \\
\midrule
\textbf{Mean} & 4.17 & 2.17 & 4.17 & 2.00 & 4.00 & 2.17 & 4.50 & 1.50 & 4.33 & 2.00 & \textbf{78.3} \\
\textbf{SD}   & 0.75 & 0.75 & 1.17 & 0.89 & 0.89 & 1.17 & 0.55 & 0.84 & 0.82 & 0.89 & \textbf{12.5} \\
\bottomrule
\end{tabular}
\label{tab:sus_items}
\end{table*}

Scores above 68 are generally considered to indicate above-average usability according to industry standards \cite{brooke_sus_2013}. Each participant's responses were scored according to the standard procedure: odd-numbered items (positively worded) were scored as the response minus 1, and even-numbered items (negatively worded) were scored as 5 minus the response. The sum of these adjusted values was then multiplied by 2.5 to yield a total SUS score per participant. Descriptive statistics revealed a mean SUS score of 78.3 (\textit{SD} = 12.5), indicating an overall positive perception of usability across participants. Scores ranged from 55 to 90, with most participants falling well above the accepted usability benchmark.

All but one participant reported SUS scores exceeding the average threshold (68), suggesting consistent user satisfaction with the system's interaction design, responsiveness, and ease of use. To further explore variability in usability perceptions, we visualized the distribution of individual item ratings using a violin plot as seen in Table \ref{tab:sus_items}. The SUS results suggest that the \textbf{\textit{PunchPulse}} demonstrates high perceived usability among its target users. Quantitatively we can conclude that the scores align with benchmarks for well-designed interactive systems and support the system's suitability for deployment in physically active, immersive environments.

\subsection{Effect of Prior Physical Activity (RQ4)}
\label{subsec:findings-vr_exp}
We explored RQ4 by analyzing how participants' baseline PA levels and prior VR exposure influenced their physiological engagement, performance pacing, and acclimation to \textbf{\textit{PunchPulse}}’s gameplay. 
Prior PA appeared most relevant not in final scores but in endurance and pacing across rounds. For instance, P3 and P5, both self-identified as highly or moderately active, maintained physically intense movement throughout the session without exhibiting signs of early fatigue. Their gameplay was marked by persistent engagement—evidenced by high duck counts and sustained punching—suggesting that baseline fitness helped buffer against the compounding intensity of each round. However, this advantage did not always translate to superior performance, indicating that physical conditioning alone did not guarantee effective interaction with the system’s spatial and timing demands. By contrast, those who reported lower exercise levels (e.g., P1 and P2) often required the full warm-up phase to calibrate their movement and orient to the task. Notably, P1 and P5 reached MVPA later in the session and displayed a more abrupt fatigue curve in the hard round, suggesting that their physiological effort was higher relative to their baseline, which refutes the theory that better prior physiological conditioning translated to higher PA levels through game engagement. Interestingly, their post-session feedback reflected a sense of personal challenge met with satisfaction, which highlights the motivational potential of progressive exertion even for less active users.

VR familiarity played a more transient but still influential role. P2, the only participant with direct prior experience using this specific system, adapted more quickly in early rounds and demonstrated efficient use of audio cues. However, they over exerted themselves in the initial rounds out of "excitement" (P2) and found it hard to sustain the same level of PA through the later rounds. Meanwhile, others with no VR background — like P4, P5 and P6 — took longer to grasp directional mechanics but were able to catch up by the end of the medium round. This suggests that while VR experience may aid initial spatial understanding, its impact diminished once participants became attuned to \textbf{\textit{PunchPulse}}’s consistent feedback structure. Warm-up requirement provided another interesting proxy for individual readiness. The three participants who declined the warm-up (P4 and P5) tended to exhibit faster starts in PA during the gameplay. This aligns with either a higher degree of confidence in spatial movement or greater baseline physical preparedness (which doesn't to align with their self-reporting of exercise levels). Conversely, those who required a warm-up benefited from the slower introduction of cues and movement, reflecting the importance of scalable onboarding into exertional gameplay.

\subsubsection{Players' Game Metrics: Scoring and Performance Patterns}
\label{subsubsec:findings-rq4-game_scores}
\begin{figure}[ht]
    \centering
    \includegraphics[width=1\linewidth]{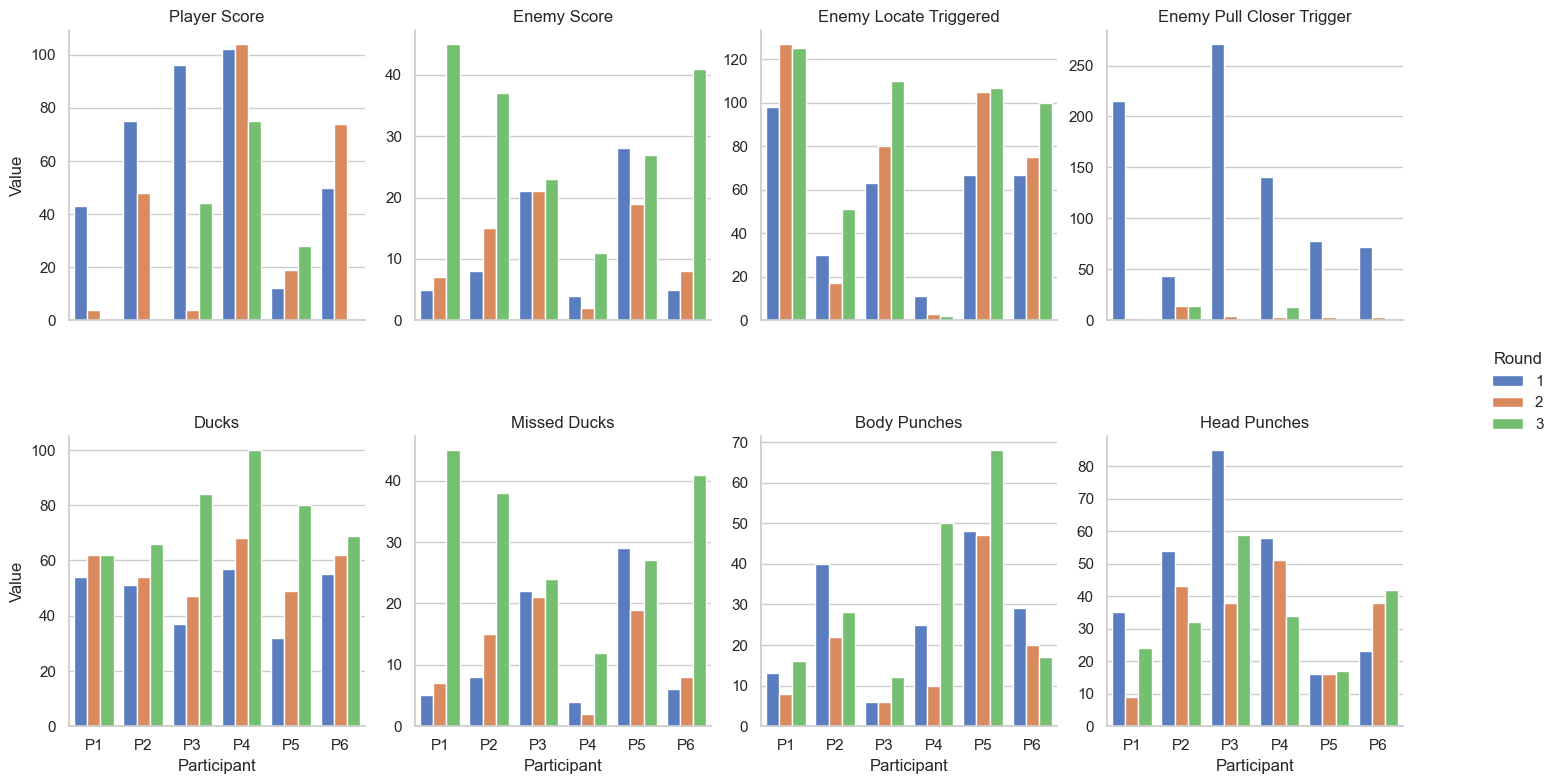}
    \caption{Participants' Game Scores and Aid Usage}
    \label{fig:player-scores}
    \Description{A grid of eight bar charts, each representing a different gameplay metric for six participants (P1 through P6) across three rounds of PunchPulse. The metrics are: Player Score, Enemy Score, Enemy Locate Triggered, Enemy Pull Closer Trigger, Ducks, Missed Ducks, Body Punches, and Head Punches. Each chart shows grouped vertical bars for each participant, color-coded by round: blue for Round 1, orange for Round 2, and green for Round 3. Most metrics show variation across rounds and participants. For example, Enemy Pull Closer Trigger has high blue bars in early rounds that drop sharply in later rounds, while Ducks and Missed Ducks generally increase over rounds. Charts are arranged in two rows and four columns for compact comparison.}
\end{figure}

We furthered our investigation of RQ4 by presenting participants' gameplay-derived performance outcomes across intensity stages. While it does not directly correlate these outcomes with prior PA or VR experience, it provides detailed insight into scoring trends, aid usage, and movement behaviors that reflect how participants performed and adapted to the increasing demands of the system.
To evaluate this engagement, we analyzed scoring trends across rounds and between participants as illustrated in Table~\ref{tab:player_scores} and Figure~\ref{fig:player-scores}. The in-game scoring system captured both offensive and defensive performance metrics, including player score, enemy score, punches landed (body and head punches), aid usage (located used - which refers to the number of times users pressed the left trigger to locate the enemy, and pull used - which refers to the number of times users pressed the right trigger to pull the enemy closer), and dodging success (ducks - which refer to the number of successful ducks, and missed ducks - which refer to the number of ducks wherein the user failed to duck and got hit by the enemy's punch). A clear pattern emerged across participants: performance generally peaked during the first or second round and dropped significantly during the third, corresponding to increased game difficulty and fatigue. P4 was the only participant (interestingly they were our only female participant as well) to maintain high scores consistently across all three rounds (102, 104, 75), demonstrating not only strong adaptability to escalating intensity but also sustained offensive execution. P4 also had among the lowest enemy scores (4, 2, 11), indicating effective defensive performance.

P3 similarly began with a high score (96) in Round 1 but exhibited a sharp drop in Round 2 (4), suggesting difficulty adapting to the sudden increase in pace and complexity. However, they partially recovered in Round 3 (44), pointing to a late-session adaptation strategy. This fluctuation mirrored behavioral observations from recordings, where P3 displayed intermittent bursts of exertion and spatial tracking lapses between rounds. P1 and P2 showed steep performance declines over time. P1’s score fell from 43 in Round 1 to 0 in Round 3, while P2 dropped from 75 to 0. In both cases, enemy scores increased proportionally in later rounds, indicating an inability to maintain defensive timing and spatial orientation under high intensity. These drops aligned with increased missed ducks and higher aid activation, suggesting a breakdown in reactive efficiency. For instance, P2’s missed ducks increased from 8 in Round 1 to 38 in Round 3, while enemy score jumped from 8 to 37.

Conversely P5, who initially performed at the lowest offensive level (score of 12 in Round 1), gradually improved across rounds (19, 28), demonstrating learning and adaptation. While their enemy scores remained high throughout, the increase in body and head punches suggests that P5 began to internalize spatial cues and target effectively by the third round. P6 followed a similar trend: scoring improved from 50 to 74 between Round 1 and 2 before dropping to 0 in the final round. However, despite the score drop, P6 maintained high punching activity throughout, with 17 head punches and 69 ducks in Round 3, indicating continued effort despite reduced accuracy or timing in offensive registration.

Inter-participant comparisons also revealed differences in play style and balance between offensive and defensive strategies. P3 and P4 consistently delivered higher head punch counts, while P1 and P2 relied more on aid usage, particularly the disabled for higher difficulty rounds "pull enemy" feature. Participants with lower aid dependence (P4 and P5) generally demonstrated greater scoring consistency across rounds, suggesting that spatially adaptive movement was more sustainable than reliance on assistive triggers. Participants who maintained or improved their scores across rounds tended to demonstrate both strategic use of feedback and resilience to escalating difficulty, whereas steep score declines were often paired with increases in missed cues, fatigue indicators, and overuse of orientation aids. These findings reinforce the importance of designing difficulty curves that scale dynamically with user performance and maintain engagement without overwhelming users' cognitive or physical resources.

\begin{table*}[ht]
\centering
\scriptsize
\caption{Player Performance Metrics Across Rounds}
\label{tab:player_scores}
\begin{tabular}{>{\centering\arraybackslash}p{1.2cm} >{\centering\arraybackslash}p{0.9cm} rrrrrrrr}
\toprule
\textbf{Player} & \textbf{Round} & \textbf{Player Score} & \textbf{Enemy Score} & \textbf{Locate Used} & \textbf{Pull Used} & \textbf{Ducks} & \textbf{Missed Ducks} & \textbf{Body Punches} & \textbf{Head Punches} \\
\midrule
\multirow{3}{*}{P1} & Round 1 & 43 & 5 & 98 & 215 & 54 & 5 & 13 & 35 \\
                    & Round 2 & 4 & 7 & 127 & 2 & 62 & 7 & 8 & 9 \\
                    & Round 3 & 0 & 45 & 125 & 1 & 62 & 45 & 16 & 24 \\
\midrule
\multirow{3}{*}{P2} & Round 1 & 75 & 8 & 30 & 43 & 51 & 8 & 40 & 54 \\
                    & Round 2 & 48 & 15 & 17 & 14 & 54 & 15 & 22 & 43 \\
                    & Round 3 & 0 & 37 & 51 & 14 & 66 & 38 & 28 & 32 \\
\midrule
\multirow{3}{*}{P3} & Round 1 & 96 & 21 & 63 & 271 & 37 & 22 & 6 & 85 \\
                    & Round 2 & 4 & 21 & 80 & 4 & 47 & 21 & 6 & 38 \\
                    & Round 3 & 44 & 23 & 110 & 1 & 84 & 24 & 12 & 59 \\
\midrule
\multirow{3}{*}{P4} & Round 1 & 102 & 4 & 11 & 141 & 57 & 4 & 25 & 58 \\
                    & Round 2 & 104 & 2 & 3 & 3 & 68 & 2 & 10 & 51 \\
                    & Round 3 & 75 & 11 & 2 & 13 & 100 & 12 & 50 & 34 \\
\midrule
\multirow{3}{*}{P5} & Round 1 & 12 & 28 & 67 & 78 & 32 & 29 & 48 & 16 \\
                    & Round 2 & 19 & 19 & 105 & 3 & 49 & 19 & 47 & 16 \\
                    & Round 3 & 28 & 27 & 107 & 0 & 80 & 27 & 68 & 17 \\
\midrule
\multirow{3}{*}{P6} & Round 1 & 50 & 5 & 67 & 72 & 55 & 6 & 29 & 23 \\
                    & Round 2 & 74 & 8 & 75 & 3 & 62 & 8 & 20 & 38 \\
                    & Round 3 & 0 & 41 & 100 & 0 & 69 & 41 & 17 & 42 \\
\bottomrule
\end{tabular}
\end{table*}
\section{Discussion}
\label{sec:discussion}

% % \hl{ }\jy{
% % Link RQs in findings. Discussion section is more about interpretation based on prior work. You need to add some scientific references to support your interpretation.
% % }

\subsection{Time-Based Gameplay Progression and MVPA Engagement}
\label{subsec:discussion-mvpa}

The design of our system intentionally emphasized time-based difficulty progression, with calibrated increases in spatial, temporal, and physical demands across gameplay rounds. The underlying objective was to scaffold users toward MVPA zones without requiring prior physical conditioning or imposing abrupt exertion thresholds. In light of this goal, the physiological data we analyzed in Section~\ref{subsection:findings-rq2_mvpa} suggests that \textbf{\textit{PunchPulse}} succeeded in not only enabling participants to reach MVPA zones, but in several cases, sustaining elevated exertion across consecutive rounds. Importantly, the sustained MVPA observed among participants cannot be solely attributed to exertion by repetition. Rather, it reflects how \textbf{\textit{PunchPulse}}’s mechanics, continuous spatial relocation of the enemy, increasingly condensed audio cues, and reduced reaction windows, effectively demanded more efficient motor responses under time pressure. These design elements did not simply add additional activity, but strategically increased challenge while preserving game flow and structure, thereby minimizing user fatigue or withdrawal. These findings support prior investigations of the influence of increasing gameplay difficulty in intensifying player engagement with the game~\cite{virtualworlds3020012, Huber_2021, Francillette_2025}. Moreover, the consistency of these outcomes across participants with diverse fitness baselines suggests that time-based scaling provided a universal pathway to MVPA, without requiring manual calibration.

From an accessibility standpoint, this outcome is significant. While conventional exercise systems often require visual confirmation~\cite{guerreiro_design_2023} to help users maintain target zones, our approach demonstrates that non-visual feedback, when integrated with a modified virtual environment -- featuring high-contrast and accessible game aids -- can effectively support BLV users in escalating gameplay intensity and maintaining performance. Notably, participants who crossed into MVPA earlier maintained higher intensity in subsequent rounds, suggesting a form of embodied training and engagement, where the body adapts to the game’s rhythm and rising difficulty~\cite{albert_effect_2022}. This presents implications and reinforcement about how temporal structuring and sensory feedback can be designed to implicitly motivate effort \cite{maculewicz_rhythmic-based_2016} specifically for BLV users, without requiring them to monitor physical metrics consciously~\cite{levi2020interpreting, bagur2025spatial}. Furthermore, \textbf{\textit{PunchPulse}}'s use of round-based structuring with clearly demarcated transitions appears to have contributed to the sustained engagement necessary for MVPA. These transitions served not just as breaks, but as cognitive reset points, moments where players could reorient and prepare for greater intensity. The deliberate pacing of these segments appears to have minimized disengagement and allowed most participants to progressively build and maintain exertion across the 60-minute session. This finding aligns with prior work emphasizing the importance of structured exercise intervals and clear phase transitions for sustaining motivation and physical activity, particularly among users with visual impairments~\cite{surakka2008motivating, guerreiro_design_2023}. By integrating these design principles, \textbf{\textit{PunchPulse}} demonstrates how accessible exergames can foster both engagement and consistent moderate-to-vigorous physical activity for BLV participants.

\subsection{Perceptions of Accessibility, Exertion, and Spatial Orientation}
\label{subsec:discussion-accessibility}
We find that participants' perceptions of the system were shaped not only by the interface itself, but by the shifting cognitive and bodily demands introduced by increasing gameplay intensity. Accessibility was not experienced as a static condition, but rather as something negotiated round by round, depending on how successfully users could integrate game feedback and regain spatial control in real time. This dynamic experience of access points to a core finding: accessibility in immersive exergames is not merely a function of affordance design, but of how well users can stabilize orientation, regulate effort, and maintain comprehension under rising cognitive load. This outcome also reinforces prior work emphasizing the importance of such aspects in crafting accessible VR experiences~\cite{yun2023fully, creed2024inclusive}.

Our behavioral data analysis results (Section~\ref{subsubsec:findings-behaviour_analysis}) suggest that spatial orientation was initially not intuitive to some but adaptive. Early hesitations, such as misinterpreting audio directions or punching in the wrong direction, reflect a need for greater anchoring in non-visual space. Yet participants often developed personal strategies for overcoming this, including verbalizing clock directions or embedding rhythmic repetition into their gameplay. These self-authored coping mechanisms are significant: they reveal that users were not passively reacting to the interface, but actively attempting to ``tune'' themselves into it. The system enabled orientation not by simplifying space, but by encouraging real-time spatial sensemaking. This reframes orientation from being a purely perceptual feature to a cognitive-motor coordination process that can be cultivated over time. As gameplay intensity increased, so did the perceptual strain. Verbal expressions of confusion, fatigue, and frustration in later rounds reflect the cognitive and physical overhead of managing both spatial and exertional feedback simultaneously. Yet notably, these frustrations did not collapse immersion. Instead, they often coincided with expressions of competitiveness, humor, and challenge-seeking. This suggests that users interpreted increased exertion not as exclusionary, but as part of the experiential value of \textbf{\textit{PunchPulse}}. The boundary between struggle and engagement was porous, reinforcing that effective responses to exertion are not inherently negative and may even enhance perceived accessibility when framed within a goal-oriented, feedback-rich environment. This interpretation is supported by the SUS results (Section~\ref{subsubsec:findings-sus}), which, despite early orientation challenges and fatigue-related commentary, indicate high overall usability. Some participants found the system easy to use in principle but still demanding in practice. This distinction underscores that usability is not synonymous with effortlessness~\cite{holzinger_2005}.
This might be because the system provided enough structure for productive failure. Users were able to fail without becoming disoriented, to exert without becoming disengaged, and to recalibrate without external intervention. These conditions signal a form of accessibility-through-resilience, where the system’s audio-haptic reinforcement and accessibility-centric aid implementation allowed participants to recalibrate dynamically as the game intensified. 

\subsection{Influence of Prior Physical Activity and VR Experience}
\label{subsec:discussion-prior-experience}
While prior PA and VR exposure influenced how players entered and adapted to the experience, these factors did not strictly determine success. Instead, they shaped the trajectory of engagement: who acclimated quickly, who needed scaffolding, and who sustained performance under pressure. We find that \textbf{\textit{PunchPulse}} did not neutralize individual differences entirely (Section~\ref{subsubsec:findings-thematic_analysis}); rather, it absorbed them, offering a structure that allowed diverse participants to perform, adapt, and exert meaningfully on their own terms (interdependently with our system). What becomes clear is that individual experience functioned more as a modulator for our system. \textbf{\textit{PunchPulse}}'s feedback structure helped flatten the learning curve by rewarding rhythmic interaction over exploratory control, allowing participants with no prior VR experience to catch up quickly and engage in meaningful PA, which was our goal. Such results align with prior work that promotes the design of digital experiences that respond to individualized user states~\cite{Houzangbe_2018, Takeshita_2022}.

\subsection{Limitations and Future Directions}
\label{subsec:limitations_and_future_directions}
While our findings provide encouraging insights into the accessibility and exertional potential of VR-based exergames for BLV users, several limitations must be acknowledged. First, the sample size was small and composed primarily of highly engaged, self-selected participants. Though the group was intentionally diverse in age, education, and prior activity levels, the limited number (N=6) restricts generalizability. These participants may not fully reflect the broader spectrum of BLV individuals, particularly those with more limited mobility, less technological access, or lower interest in PA. 

Second, although the system was designed for autonomous use, all study sessions were conducted in a semi-controlled laboratory environment with researchers present to initiate gameplay, assist with headset calibration, and ensure participant safety. While this oversight was intentionally minimal, it introduced a structured context that likely reduced friction in ways not replicable in home use. For instance, participants benefited from real-time clarification of auditory instructions, rapid troubleshooting of headset misalignment, reassurance during unfamiliar transitions and guided border patrol within the real world — all of which may have bolstered perceived usability and supported uninterrupted play. Without such scaffolding, users encountering ambiguity during calibration or gameplay may experience confusion, frustration, or disengagement. In unsupervised settings, these same points of friction, especially during onboarding, may become more pronounced. The absence of sighted assistance poses significant barriers for first-time BLV users unfamiliar with VR interfaces, as even basic setup tasks (e.g., orienting the headset, understanding menu prompts, or confirming play boundaries) depend on visual cues in most commercial systems. Furthermore, home environments introduce additional variables such as ambient noise that can mask auditory cues, spatial clutter that complicates movement, and inconsistent lighting or floor textures that disrupt spatial anchoring. These issues may compound in ways that disproportionately affect novice or low-confidence users. We propose the following solutions as future directions to improve our current implementation: (1) voice-guided setup flows, for example, could walk users step-by-step through orientation, height calibration, and boundary mapping without requiring screen-based feedback, and (2) tactile calibration tools such as textured floor markers (instead of placed white canes) or haptic anchors for directionality, could help users establish consistent starting positions and reorient themselves without relying on visual markers. These interventions would directly address usability breakdowns observed during our user test and anticipated in independent use. However, implementing them is nontrivial. Current consumer VR hardware does not support fine-grained tactile feedback natively, and most platforms lack open APIs for modifying core setup sequences. Moreover, variability in headset design, room dimensions, and user height complicates the creation of universal nonvisual setup methods. While some commercial systems offer basic audio prompts, they are typically insufficiently descriptive or responsive for users navigating VR without sight. Thus, while our design aimed for independence, our findings likely reflect an upper-bound performance shaped by the supportive lab environment. Fully autonomous home use may require not only additional interface layers but also systemic improvements in VR accessibility infrastructure. Future work must explore how such solutions can be reliably deployed across heterogeneous environments, especially for users who are new to both VR and physical activity technologies.

Third, while the study incorporated both quantitative metrics (e.g., MVPA data, SUS scores) and qualitative insights (e.g., behavioral observations and interviews), the temporal window of the study was limited to a single-session exposure. This restricts our ability to assess long-term accessibility, motivation, and physical health impact. Users’ adaptation strategies, tolerance for exertion, or enjoyment levels may evolve significantly with repeated exposure—insights which remain outside the scope of the present study. 
Finally, though the clock-based directional system provided a scalable approach to spatial orientation, not all participants found it equally intuitive. The lack of visual alignment (i.e., no physical compass-like reinforcement) occasionally led to perceptual drift. Future iterations may benefit from adaptive or personalized orientation schemes that calibrate to user preference or spatial reasoning style, especially for those newer to immersive VR. In particular, some players experienced difficulty interpreting the clock-based cues early in gameplay, often misjudging direction or reacting with delay when cues were delivered under time pressure. These issues were especially pronounced during moments of fatigue or in fast-paced sequences requiring quick repositioning. While familiarity improved over time, the learning curve suggests a need for more embodied and less abstract orientation techniques. Future design directions may explore proprioceptive calibration methods that anchor directional cues to users’ initial body posture, reducing ambiguity. Tactile overlays aforementioned, could reinforce directional feedback through touch, enabling users to recalibrate without relying solely on audio. Additionally, adaptive cueing systems that adjust timing, repetition, or phrasing based on user performance could better support spatial awareness and recovery after disorientation. These approaches offer promising paths toward improving orientation accuracy and reducing the cognitive overhead of navigation for BLV players in VR.
\section{Conclusion}
\label{sec:conclusion}
Through a combination of co-design, iterative development, and structured testing, the system demonstrated that immersive gameplay symbioses visual cues with non-visual modifications and can lead to meaningful physiological outcomes—specifically enabling users to reach and sustain MVPA zones—while remaining usable, enjoyable, and cognitively engaging. Our findings reveal that accessible exertion need not sacrifice intensity, and that orientation, autonomy, and immersion can co-exist when multisensory feedback is tightly integrated with adaptive challenge pacing. Importantly, participants' varied backgrounds in PA and VR familiarity did not preclude success; rather, the system’s design scaffolded user adaptation across those differences. As exertion increased, so did engagement—often accompanied by spontaneous strategy development, emotional investment, and satisfaction in overcoming physical and spatial challenges. These outcomes suggest that VR exergames, when designed with inclusive interaction loops, have the potential to fill critical gaps in fitness access for BLV individuals.

To further contribute to the accessible exergaming community and foster continued research and development, we have released \textit{\textbf{PunchPulse}} as an open-source project on \href{https://github.com/xability/punch-pulse}{GitHub} concurrent with this paper's publication. This release aims to support transparency, reproducibility, and broader adoption of inclusive fitness technologies.

Looking forward, this work invites further exploration of personalized exergames, long-term adherence studies, and home deployment trials to evaluate ecological validity. It also highlights the importance of shifting the conversation in accessible exercise design—from minimizing effort to supporting sustainable, meaningful intensity. In doing so, we move closer to fitness technologies that are not only inclusive, but empowering by design.

%%
%% The acknowledgments section is defined using the "acks" environment
%% (and NOT an unnumbered section). This ensures the proper
%% identification of the section in the article metadata, and the
%% consistent spelling of the heading.
\begin{acks}
  The authors wish to express their gratitude to PP1 (P2) for providing us the space to conduct our user study. We would also like to acknowledge Dr. Jonathan Freeman for allowing us to use his ITC-SOPI survey during our pilot.
\end{acks}

%%
%% The next two lines define the bibliography style to be used, and
%% the bibliography file.
\bibliographystyle{ACM-Reference-Format}
\bibliography{references/references}

\appendix

\section{Thematic Analysis}
\label{sec:thematic-analysis}

\subsection{Pilot Test Thematic Analysis}
\label{tab:pilot-thematic_analysis-table}

\twocolumn[{
\begin{center}
\scriptsize
\captionof{table}{Inductive thematic coding results from participant interviews}
\label{tab:thematic_coding}
\vspace{0.5em}
\begin{tabular}{|p{2.2cm}|p{4.5cm}|p{4.5cm}|p{4.5cm}|}
\hline
\textbf{Theme} & \textbf{Sensory Presence} & \textbf{Usability and Interaction Fluidity} & \textbf{Gamification} \\
\hline
\textbf{Spatial Awareness Through Multimodal Feedback} & 
PP1: "Where am I? I'm very concerned as to my current location. Okay. Oh, he's coming." \newline
PP1: "I had to kind of think for a moment. Like, oh yeah, how do I know where he is?" \newline
PP2: "The player was in a different direction...sometimes, when I would try to find the character and the opponent...it would send me a whole nother direction and then I'll have to orientate myself a little." &
PP1: "I just want to...buy one of these (Meta Quest 2)...Because this is the kind of workout that you could not have at home without (the HMD). You know, back in the day we had the Wii. And that was the only time. But you didn't have this much dimensionality to that because the Wii was just you facing the TV. And it was a boxing in this (front) direction. There wasn't moving, there wasn't, you know, there was ducking...I don't remember (Wii) being that intense." \newline
PP1: "So this is an alternative approach to boxing, instead of moving left right (researchers correct the perspective - it was designed to make an individual squat or physically move behind)...oh I like it." &
PP1: "I was not disoriented by the game, but rather this environment...like shadow vision, where, like I feel like myself approaching a wall. But y'all would tell me. But sometimes I was so engaged in the game that when you told me that something's going on." \newline
PP1: "It was so nice to have it (clock orientation) tell me...so all I had to do was turn around, and them I am like, okay, red humanoid - go for it...everytime I did use it, it was very helpful...except the times where that speech would overlap with another speech...not because the feature wasn't helpful, but because it was getting cut out." \\ 
\hline

\textbf{Ease of interacting with game mechanics} & 
PP2: "I just went off the (audio) cues a lot more to be honest, I kind of focused the visual part out. I was focusing a lot here (audio)...like the menus and stuff that when I went in there those were, you know...I did focus a lot more. That seems pretty well contrasted. I was listening to it (menu tts), and trying to follow what was highlighted. But...I did not try to read it." &
PP1: "I wouldn't say, it's like hard. So in the from a *** perspective. I do feel like I'm being worked out. That's for sure." \newline
PP1: "I personally don't want this to be realistic. I want this to be an active activity that I want to do, so it's fine if literally the boxer and I only have like 10 by 10, or 5 by 5 (space)." \newline
PP2: "It's a good cardio workout, but you do need...a nice area to work with." &
PP1: "Oh, he (enemy) doesn't come...(researcher: press the trigger to bring him closer) I think I got it, he comes when you press that." \newline
PP1: "(For hard level) Oh, so...it stops saying, enemy incoming. Just duck. Yeah, okay, I like that." \newline
PP1: "It doesn't have to be like an actual boxing ring...I can literally use this room. But as I approach the ropes, it's like, hey, you know, you're coming up against the ropes...which in this room would be like the wall. So that way, yes, I was fully immersed." \newline
PP1: "There was a lot of emotion involved...I think it got to a point where I wasn't even strategizing anymore." \newline
PP2: "I enjoyed this. It was really nice." \\ 
\hline

\textbf{Immersive and motivational aspects of gameplay} & 
PP1: "Am I safe?" \newline
PP1: "For me, I wasn't fully 100\% immersed. Only because I felt like I was reliant on you to tell me whether or not I was in a safe kind of...I mean, I do it would be so cool, right? if there was a room or a space that was one-to-one of what I was experiencing in the game." &
PP1: "I lost track of time...had almost 17 min...oh man yeah, I (was tired)." &
PP1: "I get a little dizzy and disoriented, because now I have 2 environments to keep track of - the virtual one, which is, I'm clearly away from the ropes in the boxing ring, and then...the real one, which is like, literally, there's a TV, there's a wall. And I don't want to hurt myself or the TV, or any of those. (Researcher: So if it was a hope, a bigger room or an open room, then you wouldn't...) No, I wanna I wanna stress this that I don't think this game should rely on a bigger room. I think it should rely on a minimum set of dimensions like, if you're gonna play this game, you need to have this amount of space available to you. What should be present is a feature within the game similar to what's available in your average VR game that allows you to define safe orders and notifies you immediately upon it does so if you would have ever come close to the boxing ring....(researchers explain the ring dynamics) I would like to try that then...I felt like...I was at a very high risk of injuring myself or property...I guess it would be nice to have defined like, Hey, I'm in this room walk the border. But yes, okay. So there's a feature that's already existing. Okay, cool." \newline
P2: "to be honest. It's a straightforward, straightforward experience." \\ 
\hline
\end{tabular}
\end{center}
}]

\subsection{User Test Thematic Analysis}
\label{subsec:usertest-thematic}

\twocolumn[{
\begin{center}
\scriptsize
\captionof{table}{Inductive thematic coding results from participant interviews}
\label{tab:thematic_coding}
\vspace{0.5em}
\begin{tabular}{|p{2.2cm}|p{4.5cm}|p{4.5cm}|p{4.5cm}|}
\hline
\textbf{Theme} & \textbf{Sensory Presence} & \textbf{Usability and Interaction Fluidity} & \textbf{Gamification} \\
\hline

Physical Engagement and Immersion &
P2: ``Sometimes I would swing and it wouldn't hit him. So I felt like at those times I was kind of detached from this virtual environment.'' \newline
P3: ``There were times where I felt like I was actually dodging a punch, like I needed to move fast or get hit.'' \newline
P4: ``I definitely felt like I was in the game world, like when he came at me, I reacted instinctively.'' 
&
P2: ``There was a few inconsistencies where he missed me and I did a duck. And there was times where he got me and I was ducked. And there was times where I would punch and I would get him. And he was far away.''\newline
P3: ``The cues were very helpful. You could hear the footsteps, but maybe if, let's say you could hear the footsteps on one ear, and like say, if it's to your right ear, the footsteps coming from the right.''\newline
P3: ``(Researcher Question - Is there anything you'd like to change in the game?) Off the top of my head, just the speed of the voice, if I were to want to play it...and I...just got to learn it once...I saw that if you keep the controllers closer together, you could learn combos more effectively. So it's really just learning the technology.'' \newline
P4: ``I played the Wii...It was fun, but we couldn't have it spinning around...here we (are able to) move too much around the room.'' \newline
P4: ``Maybe have a frame of the enemy's attacking rate, maybe change each time...I was starting to be able to gauge like one to hit as it comes forward...because I think I know when it's going to hit me. Then I hit him.''\newline
P4: ``I think someone with low vision would have felt very comfortable. Because it was like red on a grey background, otherwise (I) probably wouldn't have been able to recognize it.''\newline
P6: ``I think that I would need the support of a technical person to be able to use the system...I thought there was too much inconsistency in the system.''
&
P1: ``I wish it (enemy) did not pause as much, it waits and takes like, a pause in the first round, that makes me confused a little bit...do I move or not.''\newline
P3: ``I would recommend it (the game) to them (others)...because it's very interactive. It gets you moving more than you expect to move.''\newline
P3: ``Well, I was expecting, once you get to see the person there, or you like to get to...It's just crazy how the figure comes at you. It's cool how you can put on the headset and...you really get lost in your surroundings.''\newline
P3: ``When it's the audio thing, it's just like you play in Simon Says instead of it actually being a game.''\newline
P4: ``I thought there was too much difference (in the various difficulty levels). I would imagine that most people would find this system very fun.''\newline
P6: ``I don't know how big (the enemy) is how it's supposed to be so I feel like I move faster than the game if that makes sense...and then it recognizes that I turned and then when I started going like this (shifts to the right) then I heard the footsteps moving.'' \newline
P6: ``I didn't realize that I just like turned and started...moving. so I don't think it was registering that I was moving if that makes sense I think I needed to do it like, (take) one step, turn, wait a second, and then play. In my mind I'm like I have to turn (exactly) when I hear where he is...walk three or four steps...and then like hit.''\\
\hline

Exercise Motivation &
P1: ``I have a shoulder injury, so I workout more at home, and this seems like a nice way to sort of ease into it, I guess.''\newline
P3: ``It's a fun way of exercising because it's more like playing a game than exercising. So you're just exercising. It's just like a bonus.''\newline
P4: ``Because I was so into the game, I didn't even notice how tired I was getting until the end.''\newline
P6: ``I mean I did think...it definitely...did its job regarding exercise.''
&
P2: ``I wasn't pacing myself properly. I kind of started going to the walls in the beginning of the session. So, I think I started off relatively strong, very vigorously, if you will. And then towards the end, my energy went down, and I was exhausted.''\newline
P3: ``It was very effective on my legs...It made me sweat. I'm pretty sure it got my heart rate up.''\newline
P3: ``It's very engaging especially to those people that don't like to exercise, this is a good way of getting them active. (researcher question - Do you think it is any different than an audio game you would like to play? So does the VR actually help?) To me, yeah. Because you actually have repercussions of when you don't do the exercise, you get hit and you lose points.''
&
P2: ``This was really effective compared to last time because last time I felt like you could kind of learn how it's doing. It didn't get more difficult. It didn't feel challenging. It was a good workout last time, but the challenge factor was missing. This time, like, it was very easy to lose.''\newline
P2: ``The next best thing that could happen to this, that would be infinitely valuable, is for this to have real-time multiplayer...And they (participating players) don't have to be in the same room.''\newline
P3: ``It gets you moving more than you expect to move. It's more than just punching because it's like a full, dang near full body exercise.''\newline
P4: ``It felt like a game I'd come back to, not just a workout. I liked trying to beat my last round.''\\ \hline
\hline
\end{tabular}
\end{center}
}]

\end{document}